\documentclass[12pt,showpacs,prc]{revtex4}

\usepackage{epsfig,amssymb,amsfonts,amsmath}
\usepackage{graphicx}

\newcommand{\beq}{\begin{equation}}
\newcommand{\eeq}{\end{equation}}
\newcommand{\bea}{\begin{eqnarray}}
\newcommand{\eea}{\end{eqnarray}}
\newcommand{\beqars}[1]{\begin{eqnarray*}{#1}}
\newcommand{\eeqars}{\end{eqnarray*}}

\begin{document}
\title{Nuclear Isospin Diffusivity\\
}
\author{L.~Shi, P.~Danielewicz
}
\affiliation{ National Superconducting Cyclotron Laboratory and
 Department of Physics and Astronomy,\\
Michigan State University, East Lansing, MI 48824\\
}

\date{\today}

\begin{abstract}
The isospin diffusion and
other irreversible phenomena are discussed for a
two-component nuclear Fermi system.
The set of Boltzmann transport equations, such as employed
for reactions, are linearized, for weak deviations
of a system from uniformity, in order to arrive at nonreversible
fluxes linear in the nonuniformities.  Besides the diffusion
driven by a concentration gradient, also the diffusion driven
by temperature and pressure gradients is considered.
Diffusivity, conductivity, heat conduction and shear
viscosity coefficients are formally expressed in terms
of the responses of distribution functions
to the nonuniformities.
The linearized Boltzmann-equation set is solved, under
the approximation of constant form-factors in the distribution-function
responses, to find concrete expressions for
the transport coefficients in terms of weighted collision
integrals.   The coefficients are calculated numerically
for nuclear matter, using experimental nucleon-nucleon cross sections.
The isospin diffusivity is inversely proportional to the
neutron-proton cross section and is also sensitive to the symmetry
energy.  At low temperatures in symmetric matter, the diffusivity
is directly proportional to the symmetry energy.
\end{abstract}
 \maketitle

\section{INTRODUCTION}
\label{sec:intro}

The availability of beams largely differing in isospin content
in nuclear reactions has dramatically increased interest in
phenomena associated with the variation of that content.
In the context of peripheral reactions, this includes
interest in changes of the nuclear density profiles with the isospin content.
In central reactions, the attention has focussed, in particular, on
the dependence of isospin symmetry energy on density.  The
determination of that dependence would permit an extrapolation
of the nuclear equation of state to the neutron matter limit \cite{lat01}.
The isospin asymmetry has been used in central
reactions for projectile-target tagging in the
investigation of stopping \cite{Rami99}.

This paper deals with the irreversible transport of isospin and
other quantities in a nuclear system, as pertinent for reactions,
for small deviations from equilibrium.  In that limit, the irreversible
transport acquires universal features and is characterized in
terms of transport coefficients, that include the
isospin-diffusion coefficients.  The coefficients are derived here
for the dynamics described in terms of a Boltzmann equation set
such as used in reaction simulations \cite{Bertsch,Aichelin}.  The
main diffusion coefficient or diffusivity, characterizing isospin
diffusion driven by the gradient of asymmetry, is evaluated using
free neutron-proton cross sections.  In the past, other transport
coefficients, viscosity and heat conductivity, have been
investigated for nuclear matter
\cite{Tomonaga,gal79,PD84,hak92,hak93}. It was subsequently found
that conclusions from comparisons of reaction simulations to data
on stopping can be universally formulated in terms of the nuclear
viscosity \cite{dan02}.  It is hoped that the diffusivity can be
of such utility as that other coefficient, for the systems with a
varying isospin content.

The past studies of irreversible linear transport for nuclear
matter were primarily directed at momentum and energy. Tomonaga
\cite{Tomonaga} and Galitskii {\em et al.} \cite{gal79} obtained
the low- and high-temperature limits for the shear viscosity and
heat conductivity. Danielewicz \cite{PD84} derived results for
those coefficients valid in a wide range of nuclear densities and
temperatures. Hakim and Mornas \cite{hak93} studied different
transport coefficients within the Walecka model following the
relaxation-time approximation.

Our derivation of diffusion coefficients follows the general strategy of
Chapman and Enskog \cite{Chapman}, but here for a Fermi system, with
inclusion of mean-field effects such as appropriate for a nuclear
system.  In the next section, we discuss the diffusion coefficient
concept qualitatively and make simple estimates for the nuclear matter.
The modification of the Boltzmann equation to extend it to fermions has
been first discussed by Uhlenbeck and Uehling \cite{ueh33,ueh34}. In
Sec.\ III, we formally solve the set of Boltzmann equations for a
binary system of fermions to find thermodynamic fluxes driven by
specific thermodynamic forces and to find general but formal expressions
for the diffusion and other transport coefficients.
The transport coefficients have been (as we found) first considered for
fermions by Hellund and Uhlenbeck \cite{hel39}; compared to them, our
notation here adheres more to what is now customary for nuclear
reactions. Closely
related to the diffusivity is the electrical conductivity that is
included in our considerations. In Sec.\ IV, we obtain more specific
results for the coefficients on assuming deviations from equilibrium
suggested by the Boltzmann equation set, for specific thermodynamic
forces present. Numerical results for the coefficients are obtained in
Sec.\ V, using free NN cross sections.  We also estimate there the pace
of isospin equilibration in reactions.  We summarize our results in Sec.\
VI. More technical mathematical details and some reference information
are provided in five appendices.  In sequence, these appendices are
devoted to the definitions of macroscopic quantities, the continuity
equations, the continuity equations for an ideal fluid, the
transformations in the driving force for diffusion and to the algebra of
collision brackets.

\section{Diffusion in a Binary System}
\label{transport}

Diffusion and other irreversible transport processes occur when
a system is brought out of equilibrium.  The direction of those
processes is to bring the system back to the equilibrium.  For
small perturbations, in terms of constraints that may be set
externally, the system response is linear in the perturbation.
The coefficient of proportionality between the induced flux and
the perturbation is the transport coefficient.

In a multicomponent system with no net mass flow, irreversible
particle flows result if particle concentrations are
nonuniform.  For $N$ components, there are $N-1$ independent
flows and $N-1$ independent concentrations (since the
concentrations need to sum up to 1).  The flows are then
related to the gradients of the concentrations with an $(N-1)
\times (N-1)$ matrix of diffusion coefficients.  In a binary
system, only a single coefficient of diffusion, or diffusivity,
relates the irreversible particle flow to the nonuniformity in
concentration.  However, as we shall see, nonuniformities in
other quantities than concentration, can induce a dissipative
particle flow as well.

Our focus, obviously, is the binary system of neutrons and
protons.  However, for the sake of utility of the results
elsewhere and for the ability to examine various limits,
we shall consider a general two-component system of fermions.
An extension of those results to bosons, outside of a
condensation, will be trivial.

The two components will be denoted 1 and 2.  Then, for
the particle
$i$, the density is $n_i=N_i/{\mathcal V}$, where $N_i$ is the
particle number in some
infinitesimal volume ${\mathcal V}$.
With net density
$n=n_1+n_2$, the particle concentration for 1
is $\nu=n_1/n$ and for 2 it is $1-\nu=n_2/n$.  Moreover, with
$m_i$ representing
the mass of particle $i$, the net mass density is
$\rho = \rho_1 + \rho_2 = m_1 \, n_1 + m_2 \, n_2$, and the
mass concentration for $i$ is $c_i = m_i \, n_i/\rho$.  The
differential particle concentration
is $\delta = (n_1 -
n_2)/n$.  The different concentrations are obviously related
and thus we have $\nu=(1+\delta)/2$ and $c_1 = m_1 \,
(1+\delta)/(m_1 \, (1+\delta) + m_2 \, (1-\delta))$.  Later in
the paper, we shall primarily use the differential
concentration $\delta$ as an independent variable.

The dissipative particle flows ${\bf \Gamma}_i$ are defined relative to
the local mass velocity $\underline{\bf v}$,
\begin{equation}
{\bf \Gamma}_i = n_i \,
(\underline{\bf v}_i - \underline{\bf v}) \, ,
\label{difineflows}
\end{equation}
where $\underline{\bf v}_i$ is the local velocity of $i$'th component
and
\begin{equation}
\underline{\bf v} = (\rho_1 \, \underline{\bf v}_1 + \rho_2 \,
\underline{\bf
v}_2)/\rho \, .
\label{difinecenterofmass}
\end{equation}
We might consider
other flows such as defined relative to the local {\em particle}
velocity, but those flows are combinations of ${\bf \Gamma}_1$ and
${\bf \Gamma}_2$.  Moreover, even ${\bf \Gamma}_1$ and ${\bf \Gamma}_2$ are
redundant and we might just use ${\bf \Gamma}_1$ as an independent flow
with the flow of 2, as easily seen, given by ${\bf \Gamma}_2 = -m_1 \,
{\bf \Gamma}_1/m_2$.  Another option might be to use as independent the
differential flow defined as
\begin{equation}
{\bf \Gamma}_\delta = {\bf \Gamma}_1 -
{\bf \Gamma}_2 \, .
\label{difineIsoflow0}
\end{equation}

If the system is at uniform pressure and temperature, but there is
a small concentration gradient present, the fluxes develop
linear in the gradient, enabling us to write, e.g.\
\begin{equation}
{\bf \Gamma}_1
= -n \, D_1 \, \frac{\partial \nu}{\partial {\bf r}} \hspace*{2em}
\mbox{and} \hspace*{2em} {\bf \Gamma}_2 = - n \, D_2 \, \frac{\partial
(1-\nu)}{\partial {\bf r}} \, .
\label{diffusedefine}
\end{equation}
These are so-called Fick's laws.  Notably, the stability of an
equilibrium state requires $D_i > 0$.
Since
$m_1 \, {\bf \Gamma}_1 + m_2 \, {\bf \Gamma}_2 = 0$, we have $m_1 \, D_1 = m_2
\, D_2$.  For the differential flow, we have
\begin{equation}
{\bf \Gamma}_\delta= -
n \, D_1 \frac{\partial \nu}{\partial {\bf r}} + n \, D_2 \,
\frac{\partial (1-\nu)}{\partial {\bf r}} = - n \, D_\delta \,
\frac{\partial \delta}{\partial {\bf r}} \, .
\label{difineIsoflow}
\end{equation}
Here, the
differential coefficient is $D_\delta = (D_1 + D_2)/2$.

So far, we assumed a system at a uniform
pressure and
temperature, with just concentration changing with position.
If the variations in a system are more
complex, other nonequilibrium forces than the
concentration gradient can drive the diffusion.  This will be
explored later in the paper.  General guidance regarding the
forces which can contribute is provided by the Curie principle.  This
principle
exploits symmetry and states that the driving forces must have the same
tensor rank and parity as the flux they generate.

For the system of neutrons and protons, the differential
concentration $\delta$ becomes a concentration of the isospin and
the differential flow becomes the isospin flow, $\Gamma_\delta
\equiv \Gamma_I$.  Moreover,  the differential diffusion
coefficient becomes an isospin diffusion coefficient, $D_\delta
\equiv D_I$, and for equal masses we expect $D_I = D_p = D_n$.

It is popular to relate the concept of a diffusion coefficient
to a diffusion equation.  Indeed, if we consider a uniform system
of protons and neutrons at rest, but with the nucleon concentration changing
in space, then, from the continuity equation for the differential
density
\begin{equation}
\frac{\partial \left( n \, \delta \right)}{\partial t} =
-\nabla \cdot {\bf \Gamma}_I \, ,
\end{equation}
we get the familiar equation
\begin{equation}
\frac{\partial \delta}{\partial t} = D_I \, \nabla^2 \delta \, .
\label{eq:diffusion}
\end{equation}
Here,  for $D_I$, we have assumed a weak dependence on the concentration $\delta$.

Before turning to a derivation of rigorous results for the
diffusion and other transport coefficients, it may be instructive
to produce simple mean-free-path estimates for those coefficients.
Let us consider components of equal mass (the mass then becomes a
simple normalization coefficient in density that may be factored
out) and consider the gradient of concentration along the $x$
axis, in the medium at rest.  If we take the three coordinate
axes, then 1/6 of all particles will be primarily moving along one
of those axes in the positive or negative direction, with an
average thermal velocity $\underline{V} = \sqrt{3T/m}$, for the
distance of the order of one mean free path $\lambda$, without a
collision.  Considering the particles 1 moving through the plane
at $x=0$, they will be reflecting density at a distance $\lambda$
away.  Including the particles moving up and down through the
plane, we find for the flux
\begin{eqnarray*}
\Gamma_1 \approx \frac{1}{6}
\left(n_1 (x- \lambda) - n_1 (x + \lambda) \right) \,
\underline{V} \approx - \frac{1}{3} \lambda \, \underline{V} \,
\frac{\partial n_1}{\partial x} \, . \label{lexp}
\end{eqnarray*}
With (\ref{diffusedefine}), we then get for the diffusion
coefficient
\begin{equation}
D \sim \frac{1}{3} \lambda \, \underline{V} \sim \frac{1}{n
\, \sigma} \sqrt{\frac{T}{3 m}} \, , \label{Dest}
\end{equation}
with $\lambda \sim 1/(n \, \sigma)$. A more thorough investigation
shows that it is the cross section $\sigma_{12}$ for interaction
{\em between} the two species that enters the diffusion
coefficient.

Let us now evaluate the magnitude of the isospin
diffusion coefficient.  At temperature $T \sim 60$~MeV and
normal density $n_0 = 0.16$ fm$^{-3}$, with $\sigma_{np}
\sim 40$ mb, we find $D_I \sim 0.2$ fm$\, c$.  We will see
this to be in a rough agreement with thorough calculations.

Similarly to the above, one could employ the mean-free path arguments to
determine the better investigated coefficients: shear-viscosity
$\eta$ and heat conduction $\kappa$.  One finds $\eta \sim
\frac{1}{3} n \, m \, \underline{V} \, \lambda$ and $\kappa \sim
\frac{1}{3} n \, \underline{V} \, \lambda \, c_V$, where $c_V$
is the specific heat per particle.  For $T \sim 60$ MeV and $\sigma
\sim 40$ mb, we find $\eta \sim 30$ MeV/(fm$^2$ c) and
$\kappa \sim 0.06$ c/fm$^2$.  Up to factors, the shear
viscosity and heat conduction coefficients play the role of
diffusion coefficients in the diffusion equation for
velocity vorticity
and in the heat conduction Fourier equation identical
in form to the diffusion equation.

In the estimates above, we just considered the free motion of
particles in-between collisions.  If self-consistent mean fields
produced by the particles depend on concentration, then this
dependence, on its own, contributes to the diffusion. In the case
of nuclear matter, the interaction energy per nucleon may be well
approximated in the form quadratic in isospin asymmetry,
$E_I^{v} = a_I^{v} \, \delta^2$, where $\delta = (n_p - n_n)/n$
and $a_I^{v}$ is the interaction contribution to the asymmetry
coefficient $a_I$.  At normal density, the coefficient is $a_I^{v}
\approx 14$~MeV.  The naive expectation for two-body interactions
is that $a_I^v$ is linear in density.  At constant net density,
the quadratic dependence of the interaction energy on $\delta$
leads to the force ${\bf F}_{p,n} = \mp \frac{4 \, a_I^{v}}{n} \,
\frac{\partial n_{p}}{\partial {\bf r}}$, of opposite sign on
protons and neutrons.  The direction of the force for positive
$a_I^v$ is to reduce nonuniformity in isospin.  Under the influence
of this force, a proton accelerates for a typical time between
collisions $\Delta t = \lambda/\underline{V}$ and then, in a
collision, resets its velocity.  The described polarization effect
augments then the proton flow by
\begin{eqnarray}
\Delta {\bf \Gamma}_p = n_p \, \Delta \underline{\bf v}_p = -
\frac{4 \, a_I^{v}\, n_p}{n} \, \frac{\lambda}{2 \, m \,
\underline{V}} \frac{\partial n_{p}}{\partial {\bf r}} \, .
\label{eq:DGp}
\end{eqnarray}
In comparing with (\ref{lexp}), after
correcting for the local center of mass motion, we find that the
polarization increases the diffusion coefficient by
\begin{equation}
D_I' \sim (1 - \delta^2) \frac{a_I^{v}}{T} \,
D_I^0 \, ,
\label{eq:DI'}
\end{equation}
where $D_I^0$ represents the previous estimate in
Eq.~(\ref{Dest}).  It is apparent that the contribution of the
polarization effect is negligible for temperatures $T \gg
a_I^{v}$. However, at temperatures comparable to $a_I^{v}$, the
contribution could be significant; notably, at those temperatures Fermi
effects also need to play a role.

The isospin diffusion induced by mechanical forces
has analogy in an electric current
induced by the electric fields.
Indeed, for large enough systems, the Coulomb interactions can contribute
currents altering the concentration and, for completeness, we
evaluate the conductivity $\sigma_E$ for nuclear matter, relating the
isospin flux to the electric field,
\begin{equation}
{\bf \Gamma}_I = \sigma_E \, {\bf \mathcal E} \, ,
\end{equation}
where ${\bf \mathcal E}$ is the local electric field.

\section{Fluxes from the Boltzmann Equation Set}
\label{calculations}

\subsection{Coupled Boltzmann Equations}
\label{coupleBUU}

The two components of the binary system will be described in terms
of the quasiparticle distribution functions $f_i({\bf p}, {\bf r},t)$.
The local macroscopic quantities $h({\bf r},t)$ are expressed as momentum
integrals of $f$,
\begin{equation}
h({\bf r},t) = \frac{g}{(2\pi\hbar)^3} \int{d {\bf p}} \, { \chi ({\bf p}) f({\bf p}, {\bf r},t)}
\, ,
\label{macrodefine}
\end{equation}
where $g$ is the intrinsic degeneracy factor.  Different standard expressions for macroscopic
quantities in terms of $f$, such as pressure and heat flow, are listed in the Appendix
\ref{macroquandefine}.

The components are assumed to follow the set of coupled fermion Boltzmann
equations,
\begin{equation}
{ \frac{\partial f_i}{\partial t} } + \frac{\bf p}{m_i} \cdot
\frac{\partial{f_i}}{\partial{\bf r}} + {\bf F}_i \cdot \frac{\partial{f_i}}{\partial{\bf p}}
 = J_i \, .
\label{Boltzmann1}
\end{equation}
The terms on the l.h.s.\ account for the changes in $f_i$ due
to the movement of quasiparticles and their acceleration under the influence of mean-field and
external forces, included in ${\bf F}_i$, while the r.h.s.\
accounts for the changes in $f_i$ due to collisions.
In the following, we shall often denote the l.h.s.\ of a Boltzmann
equation as ${\mathcal D}_i$.  With $d\sigma/d\Omega$ and $v^*$ representing
the differential cross section and relative velocity, respectively, the collision
integral for particle 1 is
\begin{eqnarray}
\nonumber
J_1  = J_{11}+J_{12} & = & \frac{g}{2 (2\pi \hbar)^3} \int d^3p_{1a} \, d\Omega' \, v^*
\left( \frac{d\sigma_{11}}{d\Omega} \right) \,
\left(\tilde{f}_1 \, \tilde{f}_{1a} \,  f_{1}' \, f_{1a}'
- {f}_1 \, {f}_{1a} \,  \tilde{f}_{1}' \, \tilde{f}_{1a}' \right)\\
&& + \frac{g}{(2\pi \hbar)^3} \int d^3p_{2} \, d\Omega' \, v^*
\left( \frac{d\sigma_{12}}{d\Omega} \right) \,
\left(\tilde{f}_1 \, \tilde{f}_{2} \,  f_{1}' \, f_{2}'
- {f}_1 \, {f}_{2} \,  \tilde{f}_{1}' \, \tilde{f}_{2}' \right) \, .
\label{Boltzmann2}
\end{eqnarray}
Here, $\tilde{f} = 1-f$ is the Pauli principle factor.  The factor of $1/2$
in front of the first r.h.s.\ $J_{11}$ term, compared to the $J_{12}$ term, compensates
for the double-counting of final states when integration
is done over the full spherical angle in scattering of identical particles.
The subscript $a$ and the primes in combination with the particle subscripts
$1$ and $2$ are used to keep track of incoming and outgoing particles
for a collision.  Other than in the context of particle components,
such as here, the $1$ and $2$ subscripts will not be utilized in the paper.
The collision integral $J_2$ for particles $2$ follows from (\ref{Boltzmann2})
upon interchange of the indices $1$ and $2$.
As it stands, the set of the Boltzmann equations (\ref{Boltzmann1}),
with (\ref{Boltzmann2}),
preserves the number of each species.

In the macroscopic quantities (\ref{macrodefine}), the distribution function $f$ gets
multiplied by the degeneracy factor $g$.
When considering changes of macroscopic quantities (\ref{macrodefine}) dictated
by the Boltzmann equation (\ref{Boltzmann1}), the changing distribution function $f$
continues to be multiplied by $g$.  In the equation, the factor of $f$ for the other
particle in the collision integral $J$ is accompanied by its own factor of $g$.  As
a consequence, in the variety of physical quantities we derive, the factor of $f$ is
always accompanied by the factor of $g$, while, however, $\tilde{f}$ is not.  To
simplify the notation, in the derivations that follow, we suppress the factors of $g$,
only to restore those factors towards the end of the derivations.

When the Boltzmann equation set is used to study the temporal changes of
densities of the quantities conserved in collisions, i.e.\ number of species, energy
and momentum, local conservation laws follow.  Those conservation
laws are discussed in Appendix
\ref{continueE}.

\subsection{Strategy for Solving the Boltzmann Equation Set}
\label{approxstrategy}

Irreversible transport takes place when the system is brought out
of equilibrium such as in effect of external constraints.  Aiming
at the transport coefficients, we shall assume that the deviations
from the equilibrium are small, of the order of some parameter
$\epsilon$ that sets the scale for temporal and spatial
changes in the system.  Then the distribution functions may be
expanded in the power series in $\epsilon$ \cite{Chapman,Liboff} \begin{equation} f =
f^{(0)} + f^{(1)} + f^{(2)} + \ldots \, ,
\label{eq:approxStrategy} \end{equation} where $f^{(k)}$ represent the
consecutive terms of expansion and $f^{(0)}$ is the strict local
equilibrium solution.  The terms of expansion in $f$ may be
nominally found by expanding the collision integrals in
$\epsilon$, following (\ref{eq:approxStrategy}), expanding,
simultaneously, the derivative terms in the equations and by
demanding a consistency,
\begin{equation}
{\mathcal D}_i^{(1)} + {\mathcal
D}_i^{(2)} + \ldots = J_i^{(0)} + J_i^{(1)} + J_i^{(2)} + \ldots
\, .
\label{eq:DJ0}
\end{equation}
Here, we recognize that the derivatives,
themselves, bring in a power of $\epsilon$ into the equations and,
thus, the derivative series starts with a first order term in
$\epsilon$.

While we nominally included the zeroth-order term in the expansion of the collision
integral $J_i$, the integral vanishes for the equilibrium functions
\begin{equation}
f_j^{(0)} = \frac{1}{ \exp{\left( \frac{ \frac{({\bf p} - m_j \, {\bf v})^2}{2m_j} -\mu_j} {T} \right)} +1}
\, ,
\label{eq:feq}
\end{equation}
where $\mu_j$, ${\bf v}$ and $T$ are the local kinetic chemical
potential, velocity and temperature which are functions of ${\bf
r}$ and $t$, consistent with the Euler equations
(\ref{eq:Eulers}). Notably, the vanishing of the collision
integrals is frequently exploited in deriving the form of the
equilibrium functions, leading to the requirement that
$f_j/\tilde{f}_j$ is given by the exponential of a linear
combination of the conserved quantities. In the context of
specific transport coefficients, the boundary conditions for the
Euler equations (\ref{eq:Eulers}) may be chosen to generate just
those irreversible fluxes, and forces driving those fluxes,
that are of interest.

The equation set (\ref{eq:DJ0}) can be solved by iteration, order by order in $\epsilon$, requiring
\begin{equation}
{\mathcal D}_i^{(k)} = J_i^{(k)} \, .
\label{eq:DJ}
\end{equation}
Thus, $f_j^{(0)}$ may be introduced into ${\mathcal D}_i$, producing ${\mathcal D}_i^{(1)}$ and allowing to
find $f_j^{(1)}$.  Next, inserting $f_j^{(1)}$ into ${\mathcal D}_i$ yields ${\mathcal D}_i^{(2)}$ that
allows to find $f_j^{(2)}$ and so on.

For finding the coefficients of linear transport, only one
iteration above is necessary, since $f_i^{(1)}$, as linear in
gradients, yield dissipative fluxes that are linear in those
gradients.  The local equilibrium functions on its own produce no
dissipative fluxes, as the species local velocities ${\bf V}_j$
and heat flux ${\bf Q}$ vanish, while the kinetic pressure tensor
$\overline{\overline{P}}$ is diagonal,
\begin{subequations}
\label{eq:whole}
\begin{eqnarray}
n_i \, {\bf V}_i^{(0)} &  = & \int \frac{d^3 p}{(2 \pi \hbar)^3} \,
\frac{{\bf p}}{m_i} \, f_i^{(0)}({\bf p},{\bf r},t) = 0 \, ,\\
{\bf Q}^{(0)} & = & \sum_j  \int \frac{d^3 p}{(2 \pi \hbar)^3}\,
\frac{p^2}{2 m_j} \, \frac{{\bf p}}{m_j} \, f_j^{(0)}({\bf p},{\bf
r},t) =0 \, ,\\
\overline{\overline{P}}^{(0)} & = & \sum_j \int \frac{d^3 p}{(2 \pi
\hbar)^3}\, \frac{\overline{\overline{{\bf p} \, {\bf p}}}}{m_j}
\, f_j^{(0)}({\bf p},{\bf r},t) = \frac{2}{3} \, n \,
\underline{E}  \, \overline{\overline{1}}    \, ,
\end{eqnarray}
\end{subequations}
in the frame where the local velocity vanishes ${\bf v}({\bf r},t)=0$, with $\underline{E}$ representing
the local kinetic energy per particle.  The above fluxes
reduce the local continuity equations to the ideal-fluid Euler equations.

\subsection{Boltzmann Set in the Linear Approximation}
\label{firstorder}

We now consider the terms linear in derivatives around a given point,
i.e.\ the case of $k=1$ in (\ref{eq:DJ}), for the Boltzmann
equation set.  On representing
the distribution functions as $f_j = f_j^{(0)} + f_j^{(1)}$, we expand
the collision integrals $J_i$, to get terms $J_i^{(1)}$ linear in $f_j^{(1)}$.
Upon representing $f_j^{(1)}$ as
$f_j^{(1)} = f_j^{(0)} \, \tilde{f}_j^{(0)}\, \phi_j$, we get for the $k=1$
$i=1$
version of (\ref{eq:DJ}):
\begin{equation}
{ \frac{\partial f_1^{(0)}}{\partial t} } + \frac{\bf p}{m_1} \cdot
\frac{\partial{f_1^{(0)}}}{\partial{\bf r}} + {\bf F}_1 \cdot \frac{\partial{f_1^{(0)}}}{\partial{\bf p}}
 = - I_{11}(\phi) - I_{12}(\phi ) \, ,
\label{eq:linearized}
\end{equation}
where \begin{equation} I_{ij}(\phi) =
\frac{1}{1+\delta_{ij}} \int \frac{d^3 p_{ja}}{(2\pi)^3} \, d
\Omega' \, v^* \left( \frac{d\sigma_{ij}}{d\Omega} \right) \,
f_i^{(0)} \, f_{ja}^{(0)} \, {\tilde{f}_i^{(0)\prime}} \,
{\tilde{f}_{ja}^{(0)\prime}} \, \left(\phi_i + \phi_{ja} - \phi_i'
- \phi_{ja}' \right) \, ,
\label{eq:defineI}
\end{equation}
and where we
have utilized the property of the equilibrium functions \begin{equation}
f_i^{(0)} \, f_{ja}^{(0)} \, {\tilde{f}_i^{(0)\prime}} \,
{\tilde{f}_{ja}^{(0)\prime}} = \tilde{f}_i^{(0)} \,
\tilde{f}_{ja}^{(0)} \, {{f}_i^{(0)\prime}} \,
{{f}_{ja}^{(0)\prime}} \, . \label{fffbfb-condition} \end{equation} The
result for $i=2$, analogous to (\ref{eq:linearized}), is obtained
through an interchange of the indices 1 and 2.

The l.h.s.\ of Eq.\ (\ref{eq:linearized}) contains the derivatives
of equilibrium distribution functions with respect to $t$, ${\bf r}$ and
${\bf p}$.  These derivatives can be expressed in terms of the parameters
describing the functions (\ref{eq:feq}), i.e.\ $\mu_i$, $T$ and ${\bf v}$.
Through the use of the Euler equations (Appendix \ref{continueE}) and equilibrium
identities (Appendix \ref{FermiSysEOS}), moreover, the temporal derivatives may be
eliminated
to yield for the rescaled l.h.s.\ of (\ref{eq:linearized})
\begin{equation}
\frac{T}{f_1^{(0)} \, \tilde{f}_1^{(0)} } \, {\mathcal D}^{(1)} =
\left( \frac{p^2}{2 m_1} - \frac{5}{3}\, \underline{E} \right)
\frac{\bf p}{m_1 \, T} \cdot \frac{\partial T}{\partial {\bf r}}
+ \frac{\mathring{\overline{\overline{\bf p \, p}}}}{m_1} : \overline{\overline{\frac{\partial}{\partial
{\bf r}} {\bf v}}} + \frac{\bf p}{\rho_1} \cdot {\bf d}_{12} \, .
\label{eq:aftervariation}
\end{equation}
Here, a symmetrized traceless tensor is defined as
$\mathring{\overline{\overline{\bf x \, y}}} = \frac{1}{2} \left(
\overline{\overline{\bf x \, y}} + \overline{\overline{\bf y \, x}} \right) - \frac{1}{3}
\left({\bf x} \cdot {\bf y} \right)  \overline{\overline{1}}$, and
\begin{eqnarray}
\nonumber
{\bf d}_{12} & =  &  \frac{\rho_1 \, \rho_2}{\rho}
 \left[
\left(- \frac{{\bf F}_1}{{m_1}}+ \frac{{\bf F}_2}{{m_2}}\right)
+ T \, \frac{\partial}{\partial {\bf r}} \left( \frac{\mu_1}{m_1\, T}
- \frac{\mu_2}{m_2\, T} \right)
+  \frac{5}{3 \, T}
\left(\frac{\underline{E}_1}{m_1} - \frac{\underline{E}_2}{ m_2}\right)\,
{ \frac{\partial T}{\partial {\bf r}}}
\right] \,  , \\
& = &  \frac{\rho_1 \, \rho_2}{\rho}
 \left[
\left(- \frac{{\bf F}_1}{{m_1}}+ \frac{{\bf F}_2}{{m_2}}\right)
+ \frac{\partial}{\partial {\bf r}} \left( \frac{\mu_1}{m_1}
- \frac{\mu_2}{m_2} \right)
   +  \left( \frac{s_1}{m_1} - \frac{s_2}{m_2}    \right) \,
   \frac{\partial T}{\partial {\bf r}} \right] \, ,
\label{eq:defined12}
\end{eqnarray}
where $s_i$ is the entropy per particle
for species $i$, $s_i = (5\underline{E}_i/3 - \mu_i)/T $.
The result for species 2 in the Boltzmann equation is
obtained by interchanging the indices 1 and 2 in Eqs.\ (\ref{eq:aftervariation})
and (\ref{eq:defined12}).  Note that ${\bf d}_{21}= - {\bf d}_{12}$.

The representation (\ref{eq:aftervariation}) for the l.h.s.\ of the linearized
Boltzmann equation (\ref{eq:linearized}) exhibits the thermodynamic forces
driving the dissipative transport in a medium.  Thus, we have the tensor
of velocity gradients $\overline{\overline{\frac{\partial}{\partial {\bf r}} {\bf v}}}$
contracted in (\ref{eq:aftervariation}) with the
tensor from particle momentum.  The distortion of the momentum distribution associated
with the velocity gradients gives rise to the tensorial dissipative momentum flux in
a medium.  As to the vectorial driving forces, they all couple to the momentum in
(\ref{eq:aftervariation}) and they all can
contribute to the vector fluxes in the medium, i.e.\ the particle
and heat fluxes, as permitted by the Curie law.  The criterion that we, however, employed
in separating the driving vectors forces in (\ref{eq:aftervariation}) was that of symmetry
under the particle interchange.  When considering the diffusion in a binary system, with
the two components flowing in opposite directions in a local frame, one expects the driving
force to be of an opposite sign on the two species.  On the other hand, in the case of
the heat conduction, one expects the driving force to distort the distributions of the two species
in a similar way in the same direction.

Regarding the antisymmetric driving force in (\ref{eq:defined12}),
we may note that for conservative forces we have
\begin{equation}
{\bf F}_i = - \frac{\partial}{\partial {\bf r}} U_i \, .
\end{equation}
We can combine then the first with the second term on the r.h.s.\ of
(\ref{eq:defined12}) by introducing the net chemical potentials
$\mu_i^t = \mu_i + U_i$ and getting
\begin{equation}
{\bf d}_{12} =   \frac{\rho_1 \, \rho_2}{\rho}
 \left[
  \frac{\partial}{\partial {\bf r}} \left( \frac{\mu_1^t}{m_1}
   - \frac{\mu_2^t}{m_2} \right)
   +  \left( \frac{s_1}{m_1} - \frac{s_2}{m_2}    \right) \,
   \frac{\partial T}{\partial {\bf r}} \right] \, .
\end{equation}
For a constant temperature $T$, the driving force behind
diffusion is the gradient of difference between the chemical
potentials per unit mass, $\mu_{12}^t = \mu_1^t/m_1 -
\mu_2^t/m_2$, as expected from phenomenological considerations
\cite{Landau}. However, the temperature gradient can contribute to
the diffusion as well, which is known as the thermal diffusion or
Soret effect. We note that the vector driving forces in
(\ref{eq:aftervariation}) vanish when the temperature and
the difference of net chemical potentials per mass are uniform
throughout a system.

Given the typical constraints on a system, it can be more convenient to
obtain the driving forces in terms of the net pressure $P^t$, temperature $T$ and concentration
$\delta$, rather than $\mu_{12}^t$ and $T$.  Thus, on expressing
the potential difference as $\mu_{12}^t = \mu_{12}^t(P^t,T,\delta)$,
we get
\begin{equation}
d\mu_{12}^t = \left(\frac{\partial \mu_{12}^t}{\partial P^t}\right)_{T, \delta} \, dP^t
+ \left(\frac{\partial \mu_{12}^t}{\partial T}\right)_{P^t, \delta} \, dT
+ \left(\frac{\partial \mu_{12}^t}{\partial \delta}\right)_{P^t, T} \, d\delta \, ,
\end{equation}
and
\begin{equation}
{\bf d}_{12} = \frac{\rho_1 \, \rho_2}{ \rho} \,
    \left( \Pi_{12}^P \, {\bf \nabla} P^t + \Pi_{12}^T \, {\bf \nabla} {T}
    + \Pi_{12}^\delta \, {\bf \nabla} \delta \right) \, ,
\label{d12transformed}
\end{equation}
that we will utilize further on.  The coefficient functions are
\begin{subequations}
\begin{eqnarray}
\Pi_{12}^P & = & \left( \frac{\partial \mu_{12}^t}{\partial P^t}\right)_{T, \delta} \, , \\
\Pi_{12}^T & = &
\left(\frac{\partial \mu_{12}^t}{\partial T}\right)_{P^t, \delta}
+  \left( \frac{s_1}{m_1} - \frac{s_2}{m_2}    \right)
\, , \\
\label{eq:Pidelta} \Pi_{12}^\delta & = & \left(\frac{\partial
\mu_{12}^t}{\partial \delta}\right)_{P^t, T} \, ,
\end{eqnarray}
\end{subequations}
and specific expressions for those functions in the nuclear-matter case
are given in Appendix \ref{variable-transformation}.
Notably, however, the concentration $\delta$ may not be a convenient variable in the phase
transition region where the transformation between the chemical potential difference
and $\delta$ is generally not invertible.

With the l.h.s.\ of the linearized Boltzmann set (\ref{eq:linearized})
linear in the driving forces exhibited on the r.h.s.\ of (\ref{eq:aftervariation}), and with
the collision integrals linear in the deviation form-factors $\phi$, the form factors
need to be linear in the driving forces,
\begin{eqnarray}
\nonumber
\phi_1 & = & - {\bf A}_1 \cdot {\bf \nabla} T - \overline{\overline{B}}_1 : \mathring{\overline{\overline{\bf \nabla
 v}}} - {\bf C}_1 \cdot {\bf d}_{12} \, , \\
\phi_2 & = & - {\bf A}_2 \cdot {\bf \nabla} T - \overline{\overline{B}}_2 : \mathring{\overline{\overline{\bf \nabla
 v}}} - {\bf C}_2 \cdot {\bf d}_{12} \, ,
\label{eq:varyform}
\end{eqnarray}
where ${\bf A}$, $\overline{\overline{B}}$ and ${\bf C}$ do not depend on the forces.
On inserting (\ref{eq:varyform}) into (\ref{eq:linearized}), we get the following equations,
when keeping alternatively a selected exclusive driving force finite:
\begin{subequations}
\label{eq:diffuseconstrain}
\begin{eqnarray}
\frac{{\bf p}}{\rho_1 \, T} \, f_1^{(0)} \, \tilde{f}_1^{(0)} & = & I_{11} ({\bf C})
+ I_{12} ({\bf C}) \, , \\
-\frac{{\bf p}}{\rho_2 \, T} \, f_2^{(0)} \, \tilde{f}_2^{(0)} & = & I_{22} ({\bf C})
+ I_{21} ({\bf C}) \, ,
\end{eqnarray}
\end{subequations}
when keeping ${\bf d}_{12}$,
\begin{equation}
\frac{\mathring{\overline{\overline{\bf p
\, p}}}}{m_1 \, T} \, f_1^{(0)} \, \tilde{f}_1^{(0)} = I_{11}
(\overline{\overline{B}}) + I_{12} (\overline{\overline{B}}) \, ,
\label{eq:shearconstrain}
\end{equation}
and another one, with indices $1$
and $2$ interchanged, when keeping $\mathring{\overline{\overline{\bf
\nabla v}}}$, and, finally,
\begin{equation}
\left( \frac{p^2}{2 m} -
\frac{5}{3} \underline{E}_1 \right) \frac{\bf p }{m_1 \, T^2} \,
f_1^{(0)} \, \tilde{f}_1^{(0)} = I_{11} ({\bf A}) + I_{12} ({\bf
A}) \, ,
\label{eq:heatconstrain}
\end{equation}
and another one, with $1$
and $2$ interchanged, when keeping ${\bf \nabla} T$ (while ${\bf
d}_{12}=0$).

The linearized collision integrals $I_{ij}$ cannot change the tensorial
character of objects upon which they operate.  Moreover, the only vector that
can be locally utilized in the object construction is the momentum ${\bf p}$.
This implies, then, the following representation within the set (\ref{eq:varyform}):
\begin{subequations}
\label{variationalforms}
\begin{eqnarray}
{\bf C}_i & = & c_i (p^2) \, \frac{\bf p}{\rho_i} \,
,\label{eq:Cform}
\\
 {\bf A}_i & = & a_i (p^2)  \left( \frac{{ p}^2}{2
m_i} - \frac{5}{3} \underline{E}_i \right) \,
\frac{\bf p}{m_i \, T^2} \, , \label{eq:Aform}\\
\overline{\overline{B}} & = & b_i(p^2) \,
\mathring{\overline{\overline{{\bf p} \, {\bf p}}}} \, .
\label{eq:Bform}
\end{eqnarray}
\end{subequations}
Here, the tensorial factors are enforced by construction.
The factorization of the scalar factors is either suggested by the respective linearized
Boltzmann equation or serves convenience later on.  The unknown functions $a$,
$b$ and $c$ can be principally found by inserting (\ref{variationalforms}) into Eqs.\
(\ref{eq:diffuseconstrain})-(\ref{eq:heatconstrain}).  The resulting equations are, however,
generally quite complicated and analytic solutions are only known in some special cases.
In practical calculations, we shall contend ourselves with a power expansion for the
unknown functions.  It has been shown that any termination of the expansion will produce
lower bounds for the transport coefficients and that the lowest terms yield a predominant
contribution to the coefficients \cite{Chapman}.

\subsection{Formal Results for Transport Coefficients}
\label{formal-sol}

Before solving Eqs.\
(\ref{eq:diffuseconstrain})-(\ref{eq:heatconstrain}), we shall
obtain formal results for the transport coefficients, assuming that
solutions to (\ref{eq:diffuseconstrain})-(\ref{eq:heatconstrain})
exist. We shall start with the diffusion.  The velocity for species
$1$ is
\begin{eqnarray}
\nonumber \underline{\bf V}_1 & = & \frac{1}{n_1} \int
\frac{d^3p}{(2 \pi \hbar)^3} \, \frac{\bf p}{m_1} \, \delta f_1 =
\frac{1}{n_1} \int \frac{d^3p}{(2 \pi \hbar)^3} \, \frac{\bf
p}{m_1} \, \phi_1 \, f_1^{(0)} \, \tilde{f}_1^{(0)} \\ \nonumber &
= & T \int \frac{d^3p}{(2 \pi \hbar)^3} \, \phi_1 \, \left[ I_{11}
({\bf C}) + I_{12} ({\bf C}) \right] \\ \nonumber & = & - {\bf
\nabla} T \,  \frac{T}{3} \int \frac{d^3p}{(2 \pi \hbar)^3} \,
{\bf A}_1 \cdot \left[ I_{11} ({\bf C})
+ I_{12} ({\bf C}) \right]  \\
&& - {\bf d}_{12} \,
\frac{T}{3} \int \frac{d^3p}{(2 \pi \hbar)^3} \,
{\bf C}_1 \cdot \left[ I_{11} ({\bf C})
+ I_{12} ({\bf C}) \right] \, ,
\label{eq:variedvelocity}
\end{eqnarray}
where we have utilized (\ref{eq:varyform}) and (\ref{variationalforms}).
The contribution of a tensorial driving force to the vector flow drops out under
the integration over momentum, as required by the Curie principle.
With a result for $\underline{\bf V}_2$ analogous
to (\ref{eq:variedvelocity}), we get for the
difference of average velocities (utilized for the sake of particular
symmetry between the components)
\begin{eqnarray}
\nonumber
\underline{\bf V}_1 - \underline{\bf V}_2
& = & - {\bf \nabla} T \,  \frac{T}{3} \, \Bigg\lbrace
\int \frac{d^3p}{(2 \pi \hbar)^3} \,
{\bf A}_1 \cdot \left[ I_{11} ({\bf C})
+ I_{12} ({\bf C}) \right]  \\ \nonumber
& & + \int \frac{d^3p}{(2 \pi \hbar)^3} \,
{\bf A}_2 \cdot \left[ I_{22} ({\bf C})
+ I_{21} ({\bf C}) \right] \Bigg\rbrace \\ \nonumber
&& - {\bf d}_{12} \,
\frac{T}{3} \Bigg\lbrace \int \frac{d^3p}{(2 \pi \hbar)^3} \,
{\bf C}_1 \cdot \left[ I_{11} ({\bf C})
+ I_{12} ({\bf C}) \right]  \\ \nonumber
&& +  \int \frac{d^3p}{(2 \pi \hbar)^3} \,
{\bf C}_2 \cdot \left[ I_{22} ({\bf C})
+ I_{21} ({\bf C}) \right]
\Bigg\rbrace \\
& = & - \frac{T}{3} \left( \lbrace {\bf A},{\bf C} \rbrace \, {\bf \nabla}T +
\lbrace {\bf C},{\bf C} \rbrace \, {\bf d}_{12} \right) \, ,
\label{eq:velocitydiff}
\end{eqnarray}
where the brace product $\lbrace \cdot , \cdot \rbrace$ is an abbreviation
for the integral combinations of vectors ${\bf A}$ and ${\bf C}$, multiplying
the driving forces.  The brace product has been first introduced for a classical
gas \cite{Chapman}.  The fermion generalization of the product and
its properties are discussed in the Appendix
\ref{bracket-algebra}; see also \cite{hel39}.

The diffusion coefficient is best defined with regard to the most
common conditions under which the diffusion might occur, i.e.\ at
uniform pressure and temperature, but varying concentration.
We have then, cf.\ (\ref{difineIsoflow0}),
\begin{equation}
\underline{\bf V}_1 - \underline{\bf V}_2
= \frac{\rho}{(m_1+m_2) \, n_1 \, n_2} \, {\bf \Gamma}_\delta
= - \frac{\rho \, n \, m_{12} }{\rho_1 \, \rho_2} \, D_\delta \, {\bf \nabla} \delta \, ,
\label{eq:alterdiffusedf}
\end{equation}
where $m_{12}$ is the reduced mass, $1/m_{12}=1/m_1 + 1/m_2$.
Respectively, when $P^t$ and $T$ vary, with ${\bf d}_{12}$ given by (\ref{d12transformed}),
we write the r.h.s.\ of (\ref{eq:velocitydiff}) as
\begin{equation}
\underline{\bf V}_1 - \underline{\bf V}_2
= - \frac{\rho \, n \, m_{12} }{\rho_1 \, \rho_2} \, D_\delta \left( {\bf \nabla} \delta
+ \frac{\Pi^P}{\Pi^\delta} \, {\bf \nabla} P^t + k_T \, {\bf \nabla} T \right) \, ,
\label{eq:formaldiffuse}
\end{equation}
where, simplifying the notation, we dropped the subscripts $12$ on coefficients $\Pi$.
The diffusion coefficient in the above is given by
\begin{equation}
D_\delta = \frac{T }{3 m_{12}}\, \frac{ \Pi^\delta}{n } \, \left( \frac{\rho_1 \, \rho_2}{\rho} \right)^2
\lbrace {\bf C}, {\bf C} \rbrace \, ,
\label{eq:D12}
\end{equation}
and
\begin{equation}
k_T = \frac{\Pi^T}{\Pi^\delta} + \frac{1}{\Pi^\delta} \, \frac{\rho}{\rho_1 \, \rho_2} \,
\frac{ \lbrace {\bf A}, {\bf C} \rbrace }{\lbrace {\bf C}, {\bf C} \rbrace} \, .
\label{eq:DT}
\end{equation}

We can note that the expressions above contain $\Pi^\delta$ in
denominators.  Normally, positive nature of the derivative (\ref{eq:Pidelta}) is ensured
by the demand of the system stability.  However, across the region of a phase transition
the concentration generally changes while the chemical potentials generally
do not, so that $\Pi^\delta=0$.
While the coefficient $D_\delta$ above is the one we are after as the
standard one in describing diffusion, in the phase transition region it can be
beneficial to resort to the description of diffusion as responding to the gradient of
the potential difference in (\ref{d12transformed}).  Notably, as explained in the Appendix
\ref{bracket-algebra}, the brace product $\lbrace C, C \rbrace$ in (\ref{eq:D12})
is positive definite.  This
ensures the positive nature of $D_\delta$ away from the phase transition and, in general, ensures
that, at a constant temperature, the irreversible asymmetry flux flows in the direction from
a higher potential difference $\mu_{12}^t$ to lower.

As to the Soret effect, i.e.\
diffusion driven by the temperature gradient,
described in (\ref{eq:formaldiffuse})-(\ref{eq:DT}),
it has its
counterpart in the heat flow driven by a concentration gradient, termed
Dufour effect.  Transport coefficients for counterpart effects are
related through Onsager relations \cite{Groot} that are also borne out
by our results.  The diffusion driven by pressure is rarely of interest,
because of the usually short times for reaching the mechanical equilibrium
in a system, compared to the equilibrium with respect to temperature or
concentration.  However, an irreversible particle flux may be further
driven by external forces, such as due to an electric field ${\bf \mathcal E}$.
With the flux induced by the field given by ${\bf \Gamma}_\delta = \sigma_E \, {\bf \mathcal E}$,
where $\sigma_E$ is conductivity, with the first equality in (\ref{eq:alterdiffusedf}),
and with (\ref{eq:velocitydiff}) and
(\ref{eq:defined12}), we find for the conductivity
\begin{equation}
\sigma_E = \frac{T}{3m_{12}} \,  \left( \frac{\rho_1 \, \rho_2}{\rho} \right)^2
\, \left( \frac{q_2}{m_2} - \frac{q_1}{m_1} \right) \, \lbrace {\bf C}, {\bf C} \rbrace
= \left( \frac{q_2}{m_2} - \frac{q_1}{m_1} \right) \, \frac{n}{\Pi^\delta} \, D_\delta \, ,
\end{equation}
where $q_i$ is charge of species $i$.  We see that conductivity is closely
tied to diffusivity.

While our primary aim is to obtain coefficients characterizing the dissipative
particle transport, due to the generality of our results
we can also obtain the coefficients for the transport of energy
and momentum.  Thus, starting with the expression (\ref{eq:qdef}) in
a local frame and proceeding as in the case of (\ref{eq:variedvelocity})
and (\ref{eq:velocitydiff}), we get, with (\ref{eq:heatconstrain}),
\begin{eqnarray}
\nonumber
{\bf Q}_1 + {\bf Q}_2 & = & - \frac{T}{3} \left( \lbrace {\bf A},{\bf A} \rbrace \, {\bf \nabla}T +
\lbrace {\bf C},{\bf A} \rbrace \, {\bf d}_{12} \right)
+ \frac{5}{3} \left( \underline{E}_1 \, n_1 \, {\bf V}_1
- \underline{E}_2 \, n_2 \, {\bf V}_2 \right) \\
& = & - \frac{T}{3} \left( \lbrace {\bf A},{\bf A} \rbrace \, {\bf
\nabla}T + \lbrace {\bf C},{\bf A} \rbrace \, {\bf d}_{12} \right)
\\ \nonumber && + \frac{5}{3} \left( \frac{\underline{E}_1}{m_1} -
\frac{\underline{E}_2}{m_2} \right) \, \frac{\rho_1 \,
\rho_2}{\rho} \, \left( {\bf V}_1 - {\bf V}_2  \right) \, ,
\label{eq:formalheat2}
\end{eqnarray}
where in the second step we make use
of the condition on local velocities $\rho_1 \, {\bf V}_1 + \rho_2
\, {\bf V}_2 = 0$.  The standard procedure \cite{Landau} in coping
with the heat flux is to break it into a contribution that can be
associated with the net movement of particles and into a remnant,
driven by the temperature gradient, representing the heat conduction.
With this, the driving force ${\bf d}_{12}$ needs to be eliminated
from the heat flux in favor of the species velocities. Using
(\ref{eq:velocitydiff}), we find
\begin{eqnarray}
\nonumber {\bf Q}_1 + {\bf
Q}_2 & = & - {\bf \nabla}T \, \frac{T}{3} \left( \lbrace {\bf
A},{\bf A} \rbrace - \frac{\lbrace {\bf C},{\bf A} \rbrace ^2}
{\lbrace {\bf C},{\bf C} \rbrace} \right) \\ && + \left( {\bf V}_1
- {\bf V}_2 \right) \, \left[ \frac{5}{3} \left(
\frac{\underline{E}_1}{m_1} - \frac{\underline{E}_2}{m_2} \right)
\, \frac{\rho_1 \, \rho_2}{\rho} + \frac{\lbrace {\bf C},{\bf A}
\rbrace} {\lbrace {\bf C},{\bf C} \rbrace}   \right] \, . \end{eqnarray} The
coefficient \begin{equation} \kappa = \frac{T}{3} \left( \lbrace {\bf A},{\bf
A} \rbrace - \frac{\lbrace {\bf C},{\bf A} \rbrace ^2} {\lbrace
{\bf C},{\bf C} \rbrace} \right) \, ,
\label{eq:heatconductF}
\end{equation}
relating the heat flow to the temperature gradient, is the heat
conduction coefficient. From (\ref{eq:heatconductF}) and
considerations in Appendix \ref{bracket-algebra}, it follows that
$\kappa$ given by Eq.\ (\ref{eq:heatconductF}) is positive
definite.

The final important coefficient that we will obtain, for completeness,
is the viscosity.  The modification of the momentum flux tensor (\ref{eq:pdef}), on account
of the distortion of momentum distributions described by (\ref{eq:varyform}), is
\begin{eqnarray}
\nonumber
\overline{\overline{P}}^{(1)} & = &
\int \frac{d^3 p}{(2\pi \hbar)^3} \, \frac{\overline{\overline{\bf p \, p}}}{m_1}\,
\delta f_1 + \int \frac{d^3 p}{(2\pi \hbar)^3} \, \frac{\overline{\overline{\bf p \, p}}}{m_2}\,
\delta f_2 \\ \nonumber
& = &  - \int \frac{d^3 p}{(2\pi \hbar)^3} \, \frac{\overline{\overline{\bf p \, p}}}{m_1}\,
\left(\overline{\overline{B}}_1 : \mathring{\overline{\overline{\bf \nabla v}}}
\right) \, f_1^{(0)} \, \tilde{f}_1^{(0)}
- \int \frac{d^3 p}{(2\pi \hbar)^3} \, \frac{\overline{\overline{\bf p \, p}}}{m_2}\,
\left( \overline{\overline{B}}_1 : \mathring{\overline{\overline{\bf \nabla v}}}
 \right) \, f_2^{(0)} \, \tilde{f}_2^{(0)} \\ \nonumber
& = & - \frac{1}{5} \, \mathring{\overline{\overline{\bf \nabla v}}}
\, \left( \int \frac{d^3 p}{(2\pi \hbar)^3} \, \overline{\overline{B}}_1 :
\frac{\overline{\overline{\bf p \, p}}}{m_1}\, f_1^{(0)} \, \tilde{f}_1^{(0)}
+ \int \frac{d^3 p}{(2\pi \hbar)^3} \, \overline{\overline{B}}_2 :
\frac{\overline{\overline{\bf p \, p}}}{m_2}\, f_2^{(0)} \, \tilde{f}_2^{(0)}
\right) \\
& = & - \frac{T}{5} \, \mathring{\overline{\overline{\bf \nabla v}}} \,
\lbrace \overline{\overline{B}}, \overline{\overline{B}} \rbrace \, .
\end{eqnarray}
The coefficient of proportionality between the shear correction to the pressure
tensor and the tensor of velocity derivatives is, up to a factor of 2, the shear
viscosity coefficient
\begin{equation}
\eta = \frac{T}{10} \, \lbrace \overline{\overline{B}}, \overline{\overline{B}} \rbrace \, .
\end{equation}
As with other results for coefficients,
from Appendix \ref{bracket-algebra} it follows that the result for
$\eta$ above is positive definite, as physically required \cite{Landau}.

On account of symmetry considerations within the linear theory, the changes in temperature or concentration
do not affect the pressure tensor.  However, the situation changes if one goes beyond the linear
approximation.  For a general discussion of different higher-order effects see Ref.\ \cite{Chapman}.
As a next step, we need to find the form factors in (\ref{eq:varyform});
that requires finding the functions $a$, $b$ and $c$ in (\ref{variationalforms}) by
solving Eqs.\ (\ref{eq:diffuseconstrain})-(\ref{eq:heatconstrain}).

\section{Transport Coefficients in Terms of Cross Sections}
\label{SolveCon}

\subsection{Constraints on Deviations from Equilibrium}

Since the zeroth-order, in derivative expansion, local-equilibrium
distributions are constructed to produce the local particle densities, net velocity
and net energy, corrections to the distributions cannot alter those
macroscopic quantities.
Thus, we have locally the constraints
\begin{subequations}
\begin{eqnarray}
\delta n_i & = & \int \frac{d^3p}{(2\pi\hbar)^3} \, \delta f_i = 0 \, ,\label{eq:deltaV} \\
 \delta \left( \rho \, \underline{\bf V} \right)
& = & \int \frac{d^3p}{(2\pi\hbar)^3} \, {\bf p} \, \delta f_1
+ \int \frac{d^3p}{(2\pi\hbar)^3} \, {\bf p} \, \delta f_2 = 0 \, , \\
\delta \left( n \, \underline{E} \right)
& = & \int \frac{d^3p}{(2\pi\hbar)^3} \, \frac{ p^2}{2m_1} \, \delta f_1
+ \int \frac{d^3p}{(2\pi\hbar)^3} \, \frac{ p^2}{2m_2} \, \delta f_2 = 0 \, .
\end{eqnarray}
\end{subequations}
With driving forces being independent of each other
and with form factors in (\ref{eq:varyform}) being
independent of the forces, each of the form factors sets must separately
meet the constraints.
By inspection,
however, one can see that
the density and energy constraints are met automatically with the forms
(\ref{variationalforms}) of form factors.
Moreover, the tensorial distortion (\ref{eq:Bform}) satisfies all the
constraints.  At a general level, the ability to meet the constraints
while solving Eqs.\ (\ref{eq:diffuseconstrain})-(\ref{eq:heatconstrain})
relies on the fact that the linearized collision integrals $I_{ij}$
(in Eqs.\ (\ref{eq:linearized}) and (\ref{eq:defineI}))
nullify quantities
conserved in collisions, so a combination of the conserved quantities may be employed
in constructing the form factors $\phi_i$, ensuring that the constraints are met.
When the transport coefficients
get expressed in terms of the brace products, though,
the ensuring that the constraints are met becomes
actually irrelevant for results on the transport coefficients,
{\em because} the linearized integrals and correspondingly
brace products nullify the conserved quantities.

Given cross sections and equilibrium particle distributions,
the set of equations (\ref{eq:diffuseconstrain})-(\ref{eq:heatconstrain})
may be principally solved.  However, such a solution is generally complicated
and would likely not produce clear links between the outcome and input
to the calculations.  On the other hand, the experience has been that when
expanding the form-factor functions, $a$, $b$ and
$c$ in (\ref{variationalforms}), in power series
in $p^2$, the lowest-order results represent excellent approximations to the complete
results and are quite transparent, e.g.\ \cite{PD84}.
Thus, we adopt here the latter strategy and
test the accuracy of our results in a few selected cases.

\subsection{Diffusivity}
\label{sec:diffusivity}

If we insert (\ref{eq:Cform}) with $c_i(p^2) = c_i$ into the local velocity constraint
(\ref{eq:deltaV}), we get the requirement
\begin{equation}
\frac{c_1}{\rho_1} \int \frac{d^3p}{(2\pi\hbar)^3} \, p^2 \,
f_1^{(0)} \, \tilde{f}_1^{(0)} +
\frac{c_2}{\rho_2} \int \frac{d^3p}{(2\pi\hbar)^3} \, p^2 \,
f_2^{(0)} \, \tilde{f}_2^{(0)} = 0 \, .
\label{eq:constrain-d}
\end{equation}
After partial integrations, we find that
this is equivalent to the requirement $c_1=-c_2 \equiv c$.

When $c_i$ is constant within each species, then ${\bf C}_i$ is up
to a factor equal to momentum and, thus, gets nullified by the
linearized collision integral {\em within each species} $I_{ii}({\bf C})
= 0$. To obtain a value for $c$, we multiply the first of Eqs.\
(\ref{eq:diffuseconstrain}) by ${\bf C}_1$ and the second by ${\bf
C}_2$, add the equations side by side and integrate over momenta.
With this, we get an equation where both sides are explicitly
positive definite and, in particular, the l.h.s.\ is similar to
the l.h.s.\ of Eq.\ (\ref{eq:constrain-d}), but with an opposite
sign between the component terms.  That side of the equation can
be integrated out employing the explicit form of $f^{(0)}$ from Eq.
(\ref{eq:feq}). The other side of the resulting equation
represents $\lbrace {\bf C}, {\bf C} \rbrace$ where only the
interspecies integrals survive. On solving the equation for $c$,
we find
\begin{equation}
c = \frac{6 \, \rho_1 \, \rho_2}{\rho \, \chi_{12}}
\, ,
\end{equation}
where
\begin{equation}
\chi_{12} = g^2 \int
\frac{d^3p_1}{(2\pi\hbar)^3} \, \frac{d^3p_2}{(2\pi\hbar)^3} \,
d\Omega \, v^* \left( \frac{d\sigma_{12}}{d\Omega} \right) \, ({\bf
p}_1 - {\bf p}_1')^2 \, f_1^{(0)} \, f_2^{(0)} \,
\tilde{f}_1^{(0)\prime} \, \tilde{f}_2^{(0)\prime} \, .
\label{chidefine}
\end{equation}
The integral stems from a transformed brace
product $\lbrace {\bf C}, {\bf C} \rbrace$ and we resurrect here
the degeneracy factors $g$.  For the brace product itself, we find
\begin{equation} \lbrace {\bf C}, {\bf C} \rbrace = \frac{c^2}{2} \, \left(
\frac{\rho}{\rho_1 \, \rho_2} \right)^2 \, \chi_{12} =
\frac{18}{\chi_{12}} \, . \end{equation} On inserting this into the
diffusivity (\ref{eq:D12}), we obtain \begin{equation} D_\delta = \frac{6 T
}{m_{12}}\, \frac{ \Pi^\delta}{n  \, \chi_{12}} \, \left(
\frac{\rho_1 \, \rho_2}{\rho} \right)^2 \, . \end{equation}

In the above, we see that the diffusion coefficient depends both on
the equation of state, through the factor $\Pi^\delta$, and on the cross
section for collisions between the species, through $\chi_{12}$.  The collisions
between the species are weighted with the momentum transfer squared.  Only those
collisions between species that are characterized by large momentum transfers suppress the diffusivity
and help localize the species.  The marginalization of collisions with low
momentum transfers is a common feature of all transport coefficients.

At high temperatures the Fermi gas reduces to the Boltzmann gas.
In absence of mean-field effects, we find $\Pi^\delta \sim
\frac{2T}{m}$ for small asymmetries. The integral $\chi_{12}$ is
then of the order $n^2 \, \sigma_{12} \, p^3/m \sim n^2 \, \sigma
\, \sqrt{m\, T^3}$.  Together, these yield $D_\delta \sim
\frac{1}{n \, \sigma_{12}} \sqrt{\frac{T}{m}}$.  The precise
high-$T$ result for isotropic cross-sections in the interaction of
species with equal mass $m$ is \cite{Chapman,Liboff}
\begin{equation}
D_\delta =
\frac{3}{8n \sigma_{12}} \sqrt{\frac{T}{\pi m}} \, .
\label{eq:DBoltzmann}
\end{equation}
The square-root dependence on
temperature will be evident in our numerical results at high $T$.
With an inclusion of the mean field, with the net energy quadratic
in asymmetry, the derivative $\Pi^\delta$ gets modified into
$\Pi^\delta \sim \frac{2 \left({T} +2a_I^{v} \right)}{m}$. Thus,
the mean field enhances the diffusion.

At low temperatures, the derivative $\Pi^\delta$ is simply
proportional to the symmetry energy, $\Pi^\delta \sim \frac{4
a_I}{m}$. As to the collisional denominator of the diffusion
coefficient, at low temperatures the collisions take place only in
the immediate vicinity of the Fermi surface.  We can write the
product of equilibrium functions in the collision integral as
\begin{equation}
f_1^{(0)} \, f_2^{(0)} \, \tilde{f}_1^{(0)\prime} \,
\tilde{f}_2^{(0)\prime} = K_1 \, K_2 \, K_1' \, K_2' \, ,
\hspace*{1.5em} \mbox{where} \hspace*{1em} K_i = \frac{1}{2
\cosh{\frac{\frac{p^2}{2m_i} - \mu_i}{T}}} \, ,
\end{equation}
and, at low
$T$, $K_i \sim 2\pi \, m \, T \, \delta (p^2- p_{Fi}^2)$.  The
integration in (\ref{chidefine}) yields $\chi_{12} \sim
\sigma_{12} \, m^2 \, T^3 \, n^2/p_F^3$. In consequence, we find
that the diffusion coefficient diverges as $1/T^2$ at low
temperatures.
For the spin diffusion coefficient,
one finds
within the low temperature Landau Fermi-liquid
theory
\cite{Baym}
\begin{equation}
D_\sigma = \frac{v_F^2}{3} \, \left(1 + F_0^a \right) \,
\tau_D \, ,
\label{eq:lowT-diff-Baym}
\end{equation}
where $v_F$ is Fermi
velocity, $F_0^a$ is a spin-antisymmetric Landau coefficient and
$\tau_D$ is a characteristic relaxation time that scales as
$\tau_D \sim T^{-2}$.  The isospin diffusivity for symmetric
matter should differ from the spin diffusivity in the replacement
of the spin-antisymmetric Landau parameter with the isospin
asymmetric parameter, neither of which has a significant
temperature dependence. Thus, here consistently we find a $T^{-2}$
divergence of the diffusivity at low temperatures. Moreover, the
factor $(1+F_0^a)$ is nothing else but a rescaled symmetry energy,
with $F_0^a$ being the ratio of the interaction to the kinetic
contribution to the energy \cite{Toro}. Thus, here consistently we
find a proportionality of the diffusivity to the symmetry energy
at low temperatures.

To summarize the above results on diffusivity, we find that the diffusivity is
inversely proportional to the cross section between species for
high momentum transfers.  Moreover, whether at low or high
temperatures, the diffusivity is sensitive to the symmetry energy in the
mean-fields.  The mean-field sensitivity is
associated with the factor $\Pi^\delta = \frac{\partial
\mu_{12}}{\partial \delta} + \frac{\partial}{\partial \delta}
\left( \frac{U_1}{m_1} - \frac{U_2}{m_2} \right) = \frac{\partial
\mu_{np}}{\partial \delta} + \frac{4a_I^v}{m}$, where the last
equality pertains to the system of neutrons and protons and
$a_I^v$ represents the interaction contribution to the symmetry
energy at the relevant density.

While we obtained the diffusivity here assuming constant $c_i$ in
(\ref{eq:Cform}), we will show that the next-order term in the
expansion of $c_i$ increases the diffusion coefficient $D_\delta$
only by 2\% or less in our case of interest.

\subsection{Heat Conductivity}
\label{sec:heatco}

Evaluation of the heat conduction and shear viscosity coefficients
requires similar methodology to that utilized for the
diffusivity.  While these coefficients have been obtained
in the past for a one
component Fermi system \cite{PD84}, it can be still important to find
them for the two component system.

If we assume $a_i(p^2)=a_i$ in (\ref{eq:Aform}), then,
interestingly, we find that the momentum constraint
(\ref{eq:deltaV}) is automatically satisfied.  To obtain the
values for $a_i$, we multiply Eq.\ (\ref{eq:heatconstrain}) on both
sides by ${\bf A}_1$ and integrate over momenta and we multiply
the equation analogous to (\ref{eq:heatconstrain}) by ${\bf A}_2$
and also integrate over momenta.  As a consequence, we get a set of
equations for $a_i$ of the form
\begin{equation}
L_j = {\mathcal A}_{j1} \,
a_1 + {\mathcal A}_{j2} \, a_2 \, , \hspace*{1em} j =1,2 \, ,
\label{solconssetA}
\end{equation}
where ${\mathcal A}_{ji}$ are
coefficients independent of $a$,
\begin{equation}
{\mathcal A}_{ii} =
\frac{1}{a_{i}^2} \, \left([{\bf A},{\bf A}]_{ii} + [{\bf
A}_i,{\bf A}_i]_{12}\right)\, , \hspace*{2em} {\mathcal A}_{12}=
{\mathcal A}_{21} = \frac{1}{ a_1 \, a_2} \, [{\bf A}_1,{\bf
A}_2]_{12} \, ,
\end{equation}
cf.\ Appendix \ref{bracket-algebra}, and
\begin{equation}
L_j = \frac{1}{m_j \, T}  \left( 7\, n_j \, \underline{E_j^2} -
\frac{25}{3} \, n_j \, \left(\underline{E}_j\right)^2 \right) \, ,
\end{equation}
where $\underline{E_j^2}$ and
$\left(\underline{E}_j\right)^2$ are, respectively, the average
local square kinetic energy of species $j$ and square average
local kinetic energy of the species.

The solution to the set (\ref{solconssetA}) is
\begin{eqnarray}
\nonumber
a_1 = \left( {\mathcal A}_{22} \, L_1 - {\mathcal A}_{12} \, L_2 \right)/\Delta_{\mathcal A} \, , \\
a_2 = \left( {\mathcal A}_{11} \, L_2 - {\mathcal A}_{12} \, L_1 \right)/\Delta_{\mathcal A} \, ,
\end{eqnarray}
where the determinant is
\begin{equation}
\Delta_{\mathcal A}= {\mathcal A}_{11} \, {\mathcal A}_{22}-{\mathcal A}_{12}^2 \, .
\end{equation}
The brace product $\lbrace {\bf A} , {\bf A} \rbrace$ for use in calculating the heat conduction coefficient $\kappa$ in
(\ref{eq:heatconductF}) is
\begin{equation}
\lbrace {\bf A} , {\bf A} \rbrace = a_1 \, L_1 + a_2 \, L_2 \, .
\end{equation}
The product $\lbrace {\bf C} , {\bf A} \rbrace$ in (\ref{eq:heatconductF}) can be calculated
given the values of $a$ and $c$, and $\lbrace {\bf C} , {\bf C} \rbrace$ was already obtained
before.

\subsection{Shear Viscosity}
\label{sec:shearvi}

Evaluation of the shear viscosity coefficient $\eta$ follows
similar steps to those involved in the evaluation of $\kappa$. Thus, we
assume $b_i(p^2)=b_i$ in (\ref{eq:Bform}).  To find the
coefficient values, we convolute both sides of Eq.\
(\ref{eq:shearconstrain}) with $\mathring{\overline{\overline{\bf
p \, p}}}$ and integrate over the momenta and we do the same with
the other constraint equation for $\overline{\overline{B}}$.  The
l.h.s.\ integrations produce
\begin{equation} \frac{1}{m_i \, T} \int
\frac{d^3p}{(2\pi \, \hbar)^3} \,
\mathring{\overline{\overline{\bf p \, p}}} :
\mathring{\overline{\overline{\bf p \, p}}} \, f_i^{(0)} \,
\tilde{f}_i^{(0)} = \frac{2}{3\, m_i \, T} \int \frac{d^3p}{(2\pi
\, \hbar)^3} \, p^4 \, f_i^{(0)} \, \tilde{f}_i^{(0)} =
\frac{20}{3} \, \rho_i \, \underline{E}_i \, . \end{equation} With the
above, we get the set of equations for $b_i$: \begin{equation} \frac{20}{3} \,
\rho_j \, \underline{E}_j  = {\mathcal B}_{j1} \, b_1 + {\mathcal
B}_{j2} \, b_2 \, , \hspace*{1em} j =1,2 \, ,
\end{equation}
where the
coefficients ${\mathcal B}$ are given by,
\begin{equation} {\mathcal B}_{ii} =
[\mathring{\overline{\overline{\bf p \,
p}}},\mathring{\overline{\overline{\bf p \,
p}}}]_{ii}+[(\mathring{\overline{\overline{\bf p \,
p}}})_i,(\mathring{\overline{\overline{\bf p \, p}}})_i]_{12} \, ,
\hspace*{2em} {\mathcal B}_{12}= {\mathcal B}_{21} =
[(\mathring{\overline{\overline{\bf p \,
p}}})_1,(\mathring{\overline{\overline{\bf p \, p}}})_2]_{12} \, .
\end{equation}

Solving the set for $b$, we find
\begin{eqnarray}
\nonumber
b_1 & = & \frac{20}{3  \Delta_{\mathcal B}} \left(
\rho_1 \, \underline{E}_1 \, {\mathcal B}_{22} - \rho_2 \, \underline{E}_2 \, {\mathcal B}_{12} \right) \, , \\
b_2 & = & \frac{20}{3  \Delta_{\mathcal B}} \left(
\rho_2 \, \underline{E}_2 \, {\mathcal B}_{11} - \rho_1 \, \underline{E}_1 \, {\mathcal B}_{12} \right) \, ,
\end{eqnarray}
where the determinant is
\begin{equation}
\Delta_{\mathcal B}= {\mathcal B}_{11} \, {\mathcal B}_{22}-{\mathcal B}_{12}^2 \, .
\end{equation}
The brace product for calculating the shear viscosity coefficient $\eta = \frac{T}{10}
\lbrace \overline{\overline{B}}, \overline{\overline{B}} \rbrace$ becomes
\begin{equation}
\lbrace \overline{\overline{B}}, \overline{\overline{B}} \rbrace
= \frac{20}{3} \left( b_1 \, \rho_1 \, \underline{E}_1
+ b_2 \, \rho_2 \, \underline{E}_2
\right) \, .
\end{equation}

\section{Quantitative Results}
\label{sec:Results}

\subsection{Transport Coefficients}

We next calculate the transport coefficients as a function of density and
temperature, using experimentally measured nucleon-nucleon cross sections.
The cross sections may be altered in matter, compared to free space,
but the modifications are presumably more important at low than at the high
momentum transfers important for the transport coefficients.
With regard to the diffusivity, we first ignore any mean-field contribution
to the chemical potential difference between species.  This yields a reference
diffusivity to which the diffusivity affected by mean fields may be compared.

\begin{figure}[tbph]
\center
\includegraphics[scale=0.6]{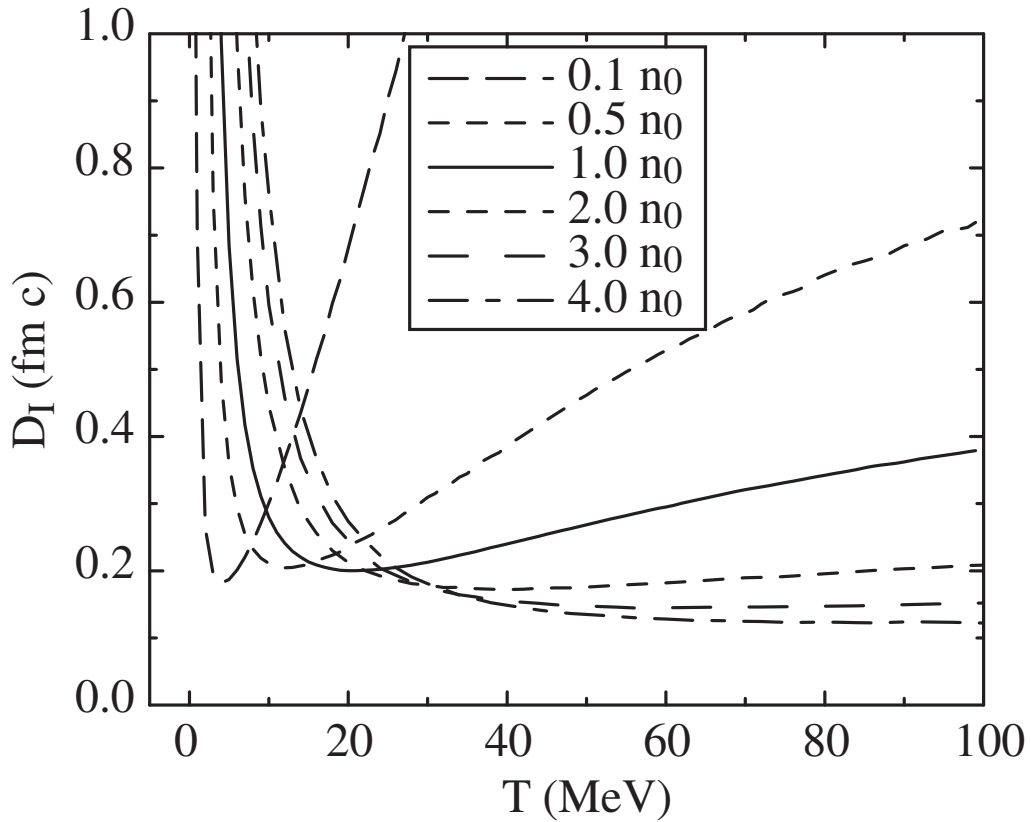}
\caption{ Isospin diffusion coefficient $D_I$ in symmetric matter,
for $U_i=0$, at different indicated densities, as a function of
the temperature $T$.  In the high-temperature limit, the diffusion
coefficient exhibits the behavior $D_I \propto \sqrt{T}/n$.
Correspondingly, at high temperatures in the figure, the largest
coefficient values are obtained for the lowest densities and the
lowest coefficient values are obtained for the highest densities.
In the low-temperature limit, the diffusion coefficient exhibits
the behavior $D_I \propto n^{3/2}/T^2$ and the order of the
results in density reverses. }
\label{diffuseFig}
\end{figure}

The diffusivity for the experimental cross sections and no
interaction contributions to the symmetry energy is shown at
$\delta=0$ and different densities $n$ in Fig.\ \ref{diffuseFig},
as a function of temperature $T$. At low temperatures, the
diffusivity diverges due to a suppression of collisions by the
Pauli principle.  At high temperatures, compared to the Fermi
energy, the role of the Pauli principle is diminished and the
diffusivity acquires a characteristic $\sqrt{T}$ dependence.  At
moderate temperatures and densities in the vicinity and above
normal, the diffusion coefficient turns out to be in the vicinity
of our original estimate of $D_I \sim 0.2$~fm$\, c$.

It should be mentioned that, for symmetric matter,
the factors for temperature and
pressure gradients in the thermodynamic force ${\bf d}_{12}$
(\ref{d12transformed}) vanish, $\Pi^P=0$ and
$\Pi^T=0$, and the brace product in (\ref{eq:DT}) vanishes,
$\lbrace{\bf A},{\bf C}\rbrace=0$, yielding
$k_T=0$ in (\ref{eq:formaldiffuse}). As physically required, the temperature
and pressure gradients produce no relative motion of neutrons and
protons for the symmetric matter.

\begin{figure}[tbph]
\center
\includegraphics[scale=0.6]{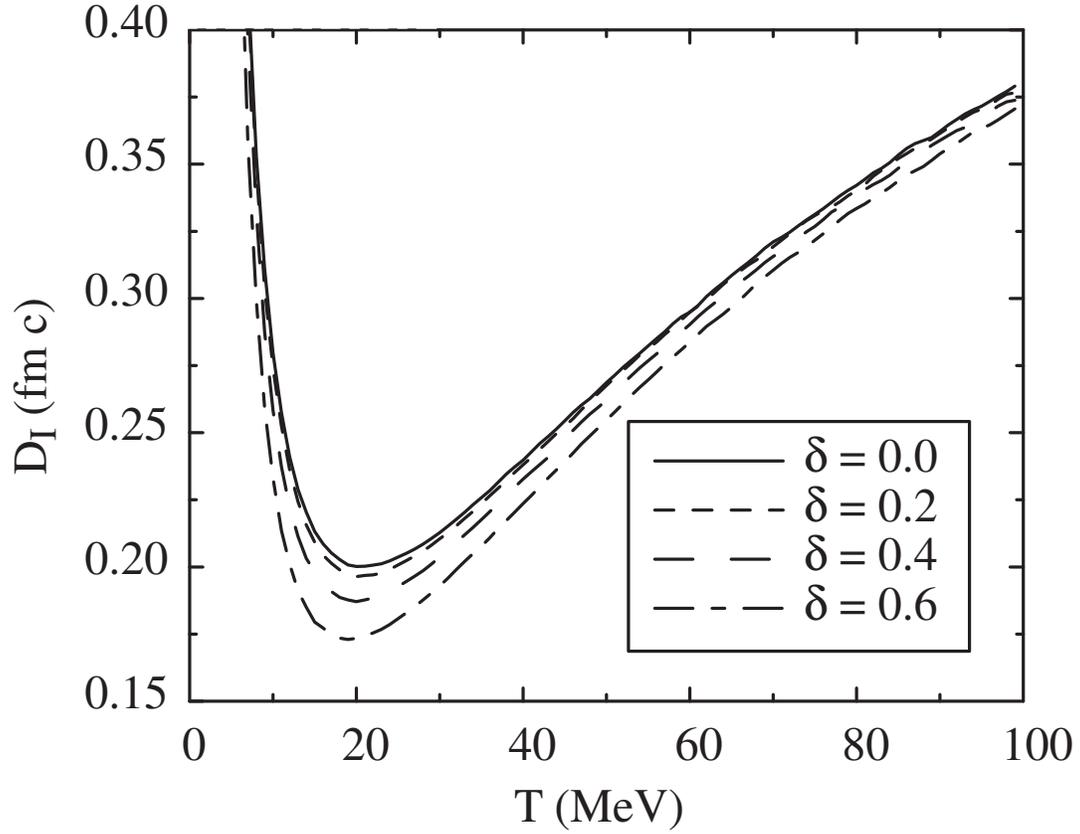}
\caption{ Isospin diffusion coefficient $D_I$ at normal density
$n=n_0=0.16\,\mbox{fm}^{-3}$ and different indicated asymmetries
$\delta$, for $U_i=0$, as a function of the temperature $T$. An
increase in the asymmetry generally causes a decrease in the
coefficient, as discussed in the text. }
\label{fig1}
\end{figure}

The diffusivity at normal density at different asymmetries is next
shown in Fig.\ \ref{fig1} as a function of temperature.  Because
of charge symmetry, the diffusivity does not depend on the sign of
$\delta$.
At low temperatures the diffusivity is generally expected to
behave as
\begin{equation}
D_I \propto \frac{\hbar^3 \, p_F^2}{m^3 \, T^2 \,
\sigma} \, ,
\label{eq:DFermi}
\end{equation}
while at high temperatures in
the manner prescribed by (\ref{eq:DBoltzmann}). With the
respective behaviors serving as a guidance, we provide a
parametrization of our numerical results for $D_\delta$ as a
function of $n$, $T$ and $\delta$,
\begin{eqnarray}
\nonumber D_I  & = &  (1 -
0.19 \, \delta^2) \, \Bigg[ \frac{11.34}{T^{2.38}} \, \left(
\frac{n}{n_0} \right)^{1.54}   \\ && + \frac{1.746}{T} \, \left(
\frac{n}{n_0} \right)^{0.56} + 0.00585 \, T^{0.913} \, \bigg(
\frac{n_0}{n} \bigg) \Bigg] \, .
\label{eq:DIFermi}
\end{eqnarray}
Here, the temperature $T$ is
in MeV and the diffusivity $D_I$ is in fm$\, c$. The
parametrization describes the numerical results to an accuracy
better than 4\% within the region of thermodynamic parameters of
$1.0 \leq n/n_0 \leq 4.0$, $10 \, \mbox{MeV} \leq T \leq 100 \,
\mbox{MeV}$ and $|\delta| \leq 0.4$.  This is, generally, the
parameter region of interest in intermediate-energy reactions.

\begin{figure}[tbph]
\center
\includegraphics[scale=0.6]{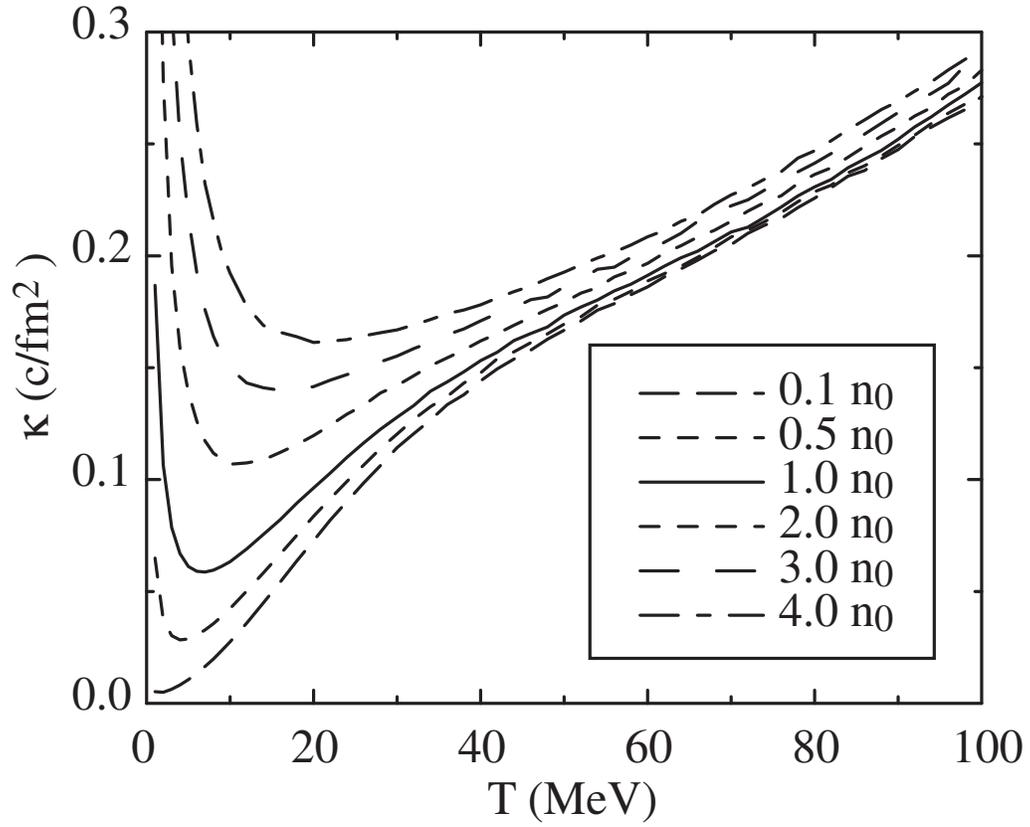}
\caption{Thermal conductivity $\kappa$ in symmetric nuclear
matter, at different indicated densities in units of $n_0$, as a
function of temperature $T$.  The conductivity increases as
density increases. } \label{thermalFig}
\end{figure}

The heat conductivity is shown for symmetric matter at different
densities in Fig.\ \ref{thermalFig}, as a function of temperature.
The results are similar to those in Ref.\ \cite{PD84},
though there the two component nature of nuclear matter was ignored
and the isospin-averaged nucleon-nucleon cross-sections have been
used.  A closer examination of results in Subsections
\ref{sec:heatco} and \ref{sec:shearvi} indicates that the use of the
isospin-averaged cross-sections is, actually, justified for
symmetric matter, when calculating the heat-conduction and
shear-viscosity coefficients.  Otherwise, however, Fig.\
\ref{thermalFig} has been based on a more complete set of cross
sections than results in \cite{PD84}. As in the case of diffusivity, the
heat conductivity diverges at low temperatures and tends to
a classical behavior at high temperatures, exhibiting there
no density dependence and being
proportional
to velocity, $\kappa \propto \sqrt{T}$.
As in the case of diffusivity,
we next provide a parametrization of our numerical results for the
heat conductivity $\kappa$ as a function of $n$, $T$ and $\delta$,
\begin{eqnarray}
\kappa & = & (1 + 0.10 \, \delta^2) \, \Bigg[
\frac{0.235}{T^{0.755}} \, \bigg( \frac{n}{n_0} \bigg)^{0.951} \nonumber\\
&& - 0.0582 \left( \frac{n}{n_0} \right)^{0.0816}
 + 0.0238 T^{0.5627} \, \left( \frac{n}{n_0} \right)^{0.0171}
\Bigg] \, .
\end{eqnarray}
Here, $T$ is again in MeV and $\kappa$ is in
c/fm$^2$.  The parameterization agrees with the numerical results
to an accuracy better than 4\% within the range of thermodynamic
parameters indicated in the case of $D_I$.

\begin{figure}[tbph]
\center
\includegraphics[scale=0.6]{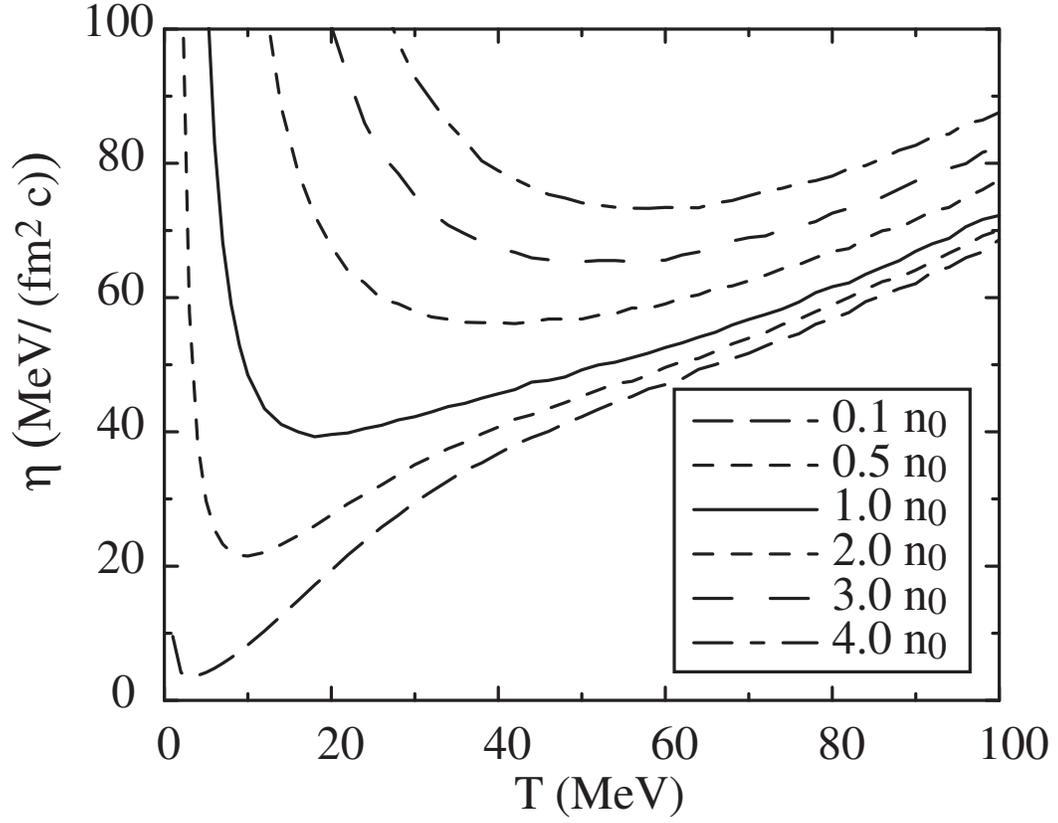}
\caption{Shear viscosity $\eta$ in symmetric nuclear matter, at
different indicated densities in units of $n_0$, as a function of
temperature $T$.  The viscosity increases as density increases. }
\label{shearFig}
\end{figure}

The shear viscosity coefficient $\eta$ is shown for symmetric matter at
at different densities, as a function of temperature, in
Fig.\ref{shearFig}.  Again, the results are similar to those in Ref.\
\cite{PD84}.  At high temperatures, the dependence on density weakens
and the viscosity becomes proportional to velocity. The numerical
results for $\eta$ are well described, to an
accuracy better than 4\% within the before-mentioned range, by
\begin{equation}
\eta  =  (1 + 0.10 \, \delta^2) \, \Bigg[
\frac{856}{T^{1.10}} \, \bigg( \frac{n}{n_0} \bigg)^{1.81}
-\frac{240.9}{T^{0.95}} \left( \frac{n}{n_0} \right)^{2.12}
 + 2.154 \, T^{0.76}
\Bigg] \, .
\label{eq:etaFermi}
\end{equation}
Here, $\eta$ is in MeV/fm$^2 \, c$ and $T$ is in
MeV.

We note in (\ref{eq:DIFermi})-(\ref{eq:etaFermi}), that the
diffusion coefficient weakly drops with increasing magnitude
of asymmetry $|\delta|$, while the
viscosity and heat conduction coefficients weakly increase.
Given the weaknesses of the dependencies, the behaviors exhibited
in parametrizations represent, in practice, averages over the
considered independent-parameter regions.  Overall, the
drop and rise in the respective coefficients with $|\delta|$ is characteristic for
a situation where the local flux of a component grows faster
than the concentration of that component.
That type of growth, with the magnitude of asymmetry,
typifies a mixture of degenerate fermion gases.
The general trends can be deduced following the mean-free-path
arguments from Sec.\ \ref{transport}.  When the average velocity rises with asymmetry,
so do the heat conduction and shear viscosity coefficients.
Additional rise for those coefficients, in the case at hand, can result
from the Pauli principle effects and from the difference between cross sections for
like and unlike particles.  Regarding the
diffusion coefficient, though, one needs to consider an irreversible part
of relative particle flux, under the condition of the concentration varying with
position.  If, starting with a given configuration of concentration gradients,
one introduces uniform changes of concentration on top,
not just the overall relative flux undergoes change but also
the reversible flux of concentration gets altered.  The rise in the relative flux
associated with the velocity of a dominant component rising with concentration is normally
more than compensated by the rise in reversible flux, leading to a reduction
in the irreversible flux and producing a reduction in diffusivity with particle
asymmetry.  A mean-field example where the reversible flux eats into the
net flux reducing the diffusivity with increasing asymmetry is the estimate in
Eq.\ (\ref{eq:DI'}), obtained there without invoking the particle Fermi statistics.

\begin{figure}[tbph]
\center
\includegraphics[scale=0.6]{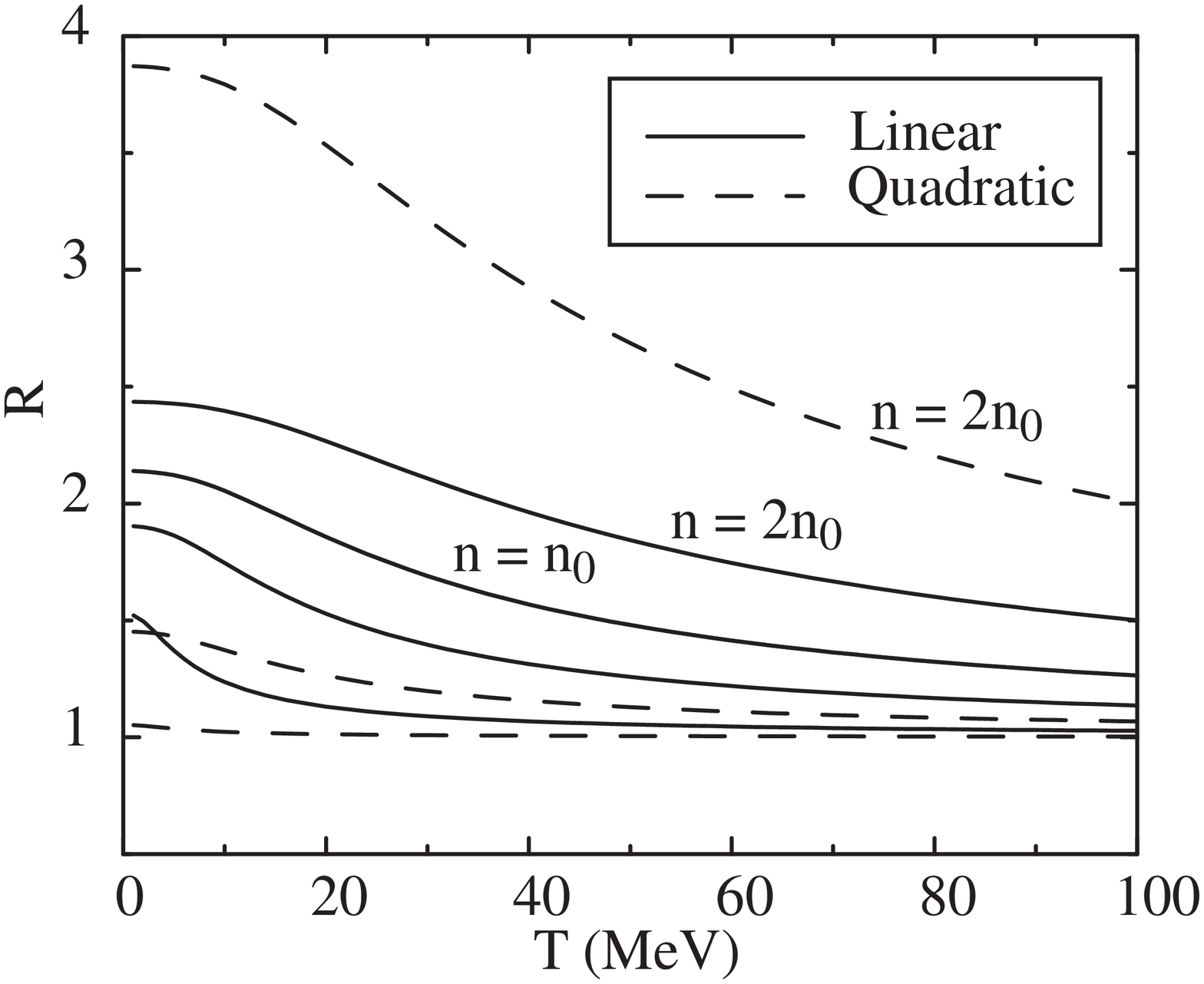}
\caption{Mean-field enhancement factor of the diffusion
coefficient in symmetric nuclear matter, $R\equiv
D_I(U_i)/D_I(U_i=0)$,
at a fixed density $n$,
as a function of temperature $T$.
The solid and dashed lines, respectively, represent the factors
for the assumed linear and quadratic dependence of the interaction
symmetry energy on density.  The lines from top to bottom are
for densities $n=2\, n_0$, $n_0$, $0.5 \, n_0$ and $0.1 \, n_0$,
respectively.  At normal density the results for the two
dependencies coincide.
} \label{ratioFig}
\end{figure}

As is found in Secs.\ \ref{firstorder} and \ref{sec:diffusivity},
the dependence of mean fields on species enters
the diffusivity through the factor $\Pi^\delta$
resulting from the variable change in thermodynamic driving force, from
the difference of chemical potentials per mass to asymmetry.
The simplest case where one can consider the impact of the mean
fields is that of the symmetric nuclear matter, at $\delta=0$.
In this case, the factor may be represented as
\begin{equation}
\Pi^\delta = \frac{1}{m} \, \left( \frac{n}{\xi}+ 4 a_I^v \right) \, ,
\end{equation}
where $\xi_i=\frac{\partial n_i }{\partial \mu_i} \equiv \xi$
(cf.\ Appendix \ref{FermiSysEOS}).  At high temperatures, we have
approximately $\xi_i \approx \frac{n_i}{T}$, so that $\frac{n}{\xi}
\approx 2T$.  The naive expectation is that $a_I^v$ has a linear
dependence on the net density,
$a_I^v= a_I^{v0}  \left( \frac{n}{n_0} \right)^\nu$, where
$a_I^{v0} = 14$~MeV and $\nu=1$.  The mean-field amplification factor
$R=\Pi^\delta(a_I^v)/\Pi^\delta(a_i^v=0)$ for the diffusion coefficient,
assuming the linear and also quadratic
density-dependence of $a_I^v$ ($\nu=1$ and 2)
is shown in Fig.\ \ref{ratioFig}.
The quadratic dependence gives higher amplification factors at $n > n_0$, than the
linear dependence, while the opposite is true at $n < n_0$.
At low temperatures and moderate to high densities
the amplification is very strong suggesting that the diffusion could
be used to probe the symmetry energy, aside from the in-medium
neutron-proton cross sections.

\subsection{Testing the Form-Factor Expansion}

The calculations of transport coefficients above have been done
assuming that the functions $a_i$, $b_i$ and $c_i$ in Eqs.\
(\ref{variationalforms}) can be approximated by constants.
In the more general case, the functions can be expanded in the
series in $p^2$, e.g.\
\begin{equation}
c_i(p^2) = c_i^{(1)} + c_i^{(2)} \, p^2
+ c_i^{(3)} \, p^4 + \ldots \,.
\end{equation}
The coefficients of the expansion
can be found by considering moments of the form-factor equations
(\ref{eq:diffuseconstrain})-(\ref{eq:heatconstrain}).  With the
more
general form of the form-factor functions, the transport coefficients
generally increase, but their rise is generally very limited.

To illustrate the magnitude of higher-order effects, we provide
in Table \ref{TablehigherOrder} results for the diffusivity
obtained in the standard first-order and in the
higher-order calculations at sample densities and temperatures.
In the indicated cases, the second-order calculations never increase
the diffusion coefficient by more than 3\% above the first-order
calculations.
The efficiency of our Monte-Carlo procedure employed to evaluate
the integrals for coefficients worsens as the order
of our calculations increases and, correspondingly, we provide only a single third-order
result for illustration.

\subsection{Isospin Equilibration}

To
test sensibility of the results on diffusivity and to
gain an elementary insight into the process of isospin equilibration in
a reaction,
we carry out a schematic consideration of the equilibration.  For
definiteness, and to ensure a level of applicability for our
consideration, we take the case of a
$^{96}$Ra + $^{96}$Zr reaction at $E_{lab}/A=100\, \mbox{MeV}$.
Densities in the central region of the reaction are
not far from normal.  Following the degenerate Fermi-gas limit,
the temperature in the central region can be estimated with 
$T \sim \sqrt {E_{lab}/(2a)} \sim
20 \, \mbox{MeV}$ using $a \approx A/(8 \, \mbox{MeV})$; deviations
from the degenerate limit yield a bit higher value.  Under
those conditions, basing on Figs.\ \ref{diffuseFig}
and \ref{ratioFig}, we estimate the streaming contribution to the diffusivity
in the central region
at $0.21\, \mbox{fm}\, c$ and the mean-field contribution at
$0.20\, \mbox{fm}\, c$, for a net $D_I \approx 0.41 \, \mbox{fm}\, c$.

Considering the direction perpendicular to the plane of contact
between the nuclei, with nuclei extending a distance $L \sim
(A/n_0)^{1/3} \sim 8 \, \mbox{fm}$ both ways from the interface,
we may use the one-dimensional diffusion equation to estimate the
isospin equilibration \begin{equation} \frac{\partial \delta}{\partial t} =
D_I \, \frac{\partial^2 \delta}{\partial x^2} \, ,
\label{eq:1Ddiffusion} \end{equation} where $x$ is the direction
perpendicular to the interface, cf.\ Eq.\ (\ref{eq:diffusion}).
With isospin flux vanishing at the boundaries of the region
$[-L,L]$, the solution to (\ref{eq:1Ddiffusion}) is
\begin{equation}
\delta(x,t) = \delta_\infty + \sum_{n=1}^\infty a_n \, \sin{k_n \,
x} \, \exp{(-D_I \, k_n^2 \, t)} + \sum_{n=1}^\infty b_n \, \cos{q_n
\, x} \, \exp{(-D_I \, q_n^2 \, t)} \,
\label{eq:deltaxt}
\end{equation}
where $k_n \, L = \left(n -
\frac{1}{2} \right) \, \pi$ and $q_n \, L = n \, \pi$.  The
coefficients $a_n$ and $b_n$ are determined by the initial
conditions and, in the case in question, $b_n=0$.

The different terms in the expansion (\ref{eq:deltaxt}) correspond to the
different levels of detail in the distribution of concentration,
as characterized by the different wavevectors.  We see that the greater
the detail the faster the information is erased,
with the erasure rates proportional to wavevectors squared.
The overall
distribution tends towards $\delta_\infty$ as $t\rightarrow \infty$.
The late-stage approach to equilibrium is governed by the rate
for the term with the lowest wavevector, i.e.~$a_1$.  Defining
the isospin equilibration time $t_H$ as one for which the original isospin
asymmetry between the nuclei is reduced by half, we get from
(\ref{eq:deltaxt}) an estimate for the reaction
\begin{equation}
t_H \approx \frac{\ln{2}}{D_I \, k_1^2} = \frac{4 \ln{2} \, L^2}{\pi^2 \, D_I}
\sim 44 \, \mbox{fm}/c \, ,
\end{equation}
for the case above.
When we carry out the full respective Boltzmann-equation simulations of the $100\,
\mbox{MeV/nucleon}$
$^{96}$Ru + $^{96}$Zr reactions, at the impact
parameter of $b=5 \, \mbox{fm} \gtrsim L/2$ (to ensure a neck comparable to that in
the consideration),
we find that, indeed, the nuclei need to be in contact for about $40 \, \mbox{fm}/c$
for the isospin asymmetry to drop to the half of original value.

\section{Summary}
\label{sec:Summary}

Diffusion and other irreversible transport phenomena
have been discussed for a binary Fermi system
close to equilibrium.
For weak nonuniformities, the irreversible fluxes
are linear in the uniformities, with the characteristic
transport proportionality-coefficients dependent only
on the equilibrium system.  It is hoped that, in an analogy to
how the nuclear equation of state and symmetry energy
are employed, the coefficient of diffusion
can be employed to characterize
reacting nuclear systems with respect to isospin transport.

Following a qualitative discussion of the irreversible transport
in the paper, the set of coupled Boltzmann-Uhlenbeck-Uehling equations was
considered for a binary system, assuming slow macroscopic temporal and spatial changes.
The slow changes allow to solve the equation set by iteration, with
the lowest-order solution being the local equilibrium
distributions.  In the next order, corrections to those distributions were
obtained, linear in the thermodynamic driving forces associated
with the system nonuniformities.  These corrections produce
irreversible fluxes linear in the forces.  The transport coefficients
have been formally expressed in terms of brace products of the responses of
distribution functions to the driving forces.
The considered coefficients include diffusivity, conductivity,
heat conduction and shear viscosity.

The set of the linearized Boltzmann equations
was, further, explicitly solved under the assumption
of simplified distribution-function responses
to the thermodynamic driving forces.
The solutions to the equations led to explicit expressions for the
transport coefficients,
with the diffusivity given in terms of the collision
integral for collisions between the two species
weighted by the momentum transfer
squared.  Besides associated sensitivity to the cross section for collisions
between the species, the diffusivity is also sensitive to the dependence of
mean fields on the species.
The collisions {\em between the species} are those that inhibit the relative motion of
the species; the difference between mean fields affects the relative acceleration
and, in combination with the collisions, the stationary diffusive flux that is
established.

We calculated the isospin diffusivity for nuclear matter,
using experimental nucleon-nucleon cross
sections for species-independent mean-fields.  At low temperatures
and high densities, the diffusivity diverges due a suppression of
collisions by the Pauli principle.  At high temperatures, the diffusivity
is roughly proportional to the average velocity and is inversely proportional
to the density.
The diffusivity weakly decreases with an increase
in the absolute magnitude of asymmetry.  We provided
an analytic fit to our numerical results.  For completeness, we also
calculated the heat conduction and shear viscosity coefficients and
provided fits to those.  Moreover, we calculated the diffuseness mean-field
enhancement factor for symmetric matter, assuming a couple of dependencies of
the symmetry energy on density.  At low temperatures, the enhancement
factor is simply proportional to the net symmetry energy divided
by the kinetic symmetry energy.  Considering the expansion of the
form-factors in distribution-function responses, we demonstrated
that corrections to the Boltzmann-equation transport coefficients,
beyond the approximations we employed, are small.
Finally, we produced an elementary
estimate for isospin equilibration in a low impact-parameter collision.

\begin{acknowledgments}
The authors thank Betty Tsang and Bill Lynch
for the collaboration on a related
parallel subject.  This work was supported by the National Science
Foundation under the Grant PHY-0070818.

\end{acknowledgments}

\appendix
\section{Macroscopic Quantities}
\label{macroquandefine}

We shall consider different types of macroscopic quantities, either net or for
separate components, either in the general frame of observation or in a local
frame.  For a single component $i$ in the
observation frame, the density $n_i$, mean velocity
$\underline{\bf v}_i$,
mean kinetic energy
$\underline{e_i}$, momentum flux tensor $\overline{\overline{p}}_i$ and kinetic
energy flux ${\bf q}_i$, are given in terms of the
distribution $f_i$, respectively, as
\begin{subequations}
\label{averagedefine}
\begin{eqnarray}
n_i ({\bf r},t) & = & \frac{g}{(2 \pi \hbar)^3} \int {d^3 p} \, f_i({\bf p},{\bf r},t) \, , \\
n_i \, \underline{\bf v}_i & = & \frac{g}{(2 \pi \hbar)^3}
\int {d^3 p}\, \frac{\bf p}{m_i} \, f_i({\bf p},{\bf r},t) \, , \\
n_i \, \underline{e}_i & = &  \frac{g}{(2 \pi \hbar)^3} \int {d^3
p}\, \frac{p^2}{2 m_i} \, f_i({\bf p},{\bf r},t) \, , \\
\label{eq:pdef} \overline{\overline{p}}_i & = &  \frac{g}{(2 \pi
\hbar)^3}
\int {d^3 p}\, \frac{\overline{\overline{{\bf p \,  p}}}}{m_i} \, f_i({\bf p},{\bf r},t) \, , \\
{\bf q}_i & = &  \frac{g}{(2 \pi \hbar)^3} \int {d^3 p}\,
\frac{p^2}{2 m_i} \, \frac{{\bf p}}{m_i} \, f_i({\bf p},{\bf r},t)
\, . \label{eq:qdef}
\end{eqnarray}
\end{subequations}

The net quantities result from combining the component
contributions.  Thus, the net density is $n=n_1 + n_2$, the
net mass density is $\rho =\rho_1+\rho_2= m_1 \, n_1 + m_2 \, n_2$
while the net velocity $\underline{\bf v}$ is obtained from $\rho \,
\underline{\bf v} = \rho_1 \, \underline{\bf v}_1 + \rho_2 \,
\underline{\bf v}_2$. The kinetic energy $\underline{e}$ averaged
over all particles is given by $n\, \underline{e} = n_1 \,
\underline{e}_1 + n_2 \, \underline{e}_2$, the net momentum flux
is $\overline{\overline{p}}= \overline{\overline{p}}_1 +
\overline{\overline{p}}_2$ and the net kinetic energy flux is
${\bf q} = {\bf q}_1 + {\bf q}_2$.

Local quantities are those calculated with momenta transformed to the local mass frame, i.e.\
following the substitution
${\bf p} \rightarrow  {\bf p} - m_i \, \underline{\bf v}$.  To distinguish local
quantities from those in the observation frame, when the frame matters, the local quantities
will be capitalized.  The local momentum flux tensor $\overline{\overline{P}}$ is the kinetic
pressure tensor and the local kinetic energy flux ${\bf Q}$ is the heat flux.

\section{continuity equations}
\label{continueE}

The collisions in the Boltzmann equation
set (\ref{Boltzmann1}) conserve the quasiparticle
momentum and energy and the species identity.
This leads to local conservation laws for the corresponding
macroscopic quantities.

Let $\chi_j ({\bf p})$ represent one of the quasiparticle
quantities conserved in collisions, $\chi_j ({\bf p}) =
\delta_{ij}$, ${\bf p}$ or $p^2/2m_j$.  For those quantities,
the integration with collision integrals produces
\begin{equation}
\sum_j \int d^3 p
\, \chi_j \, J_j = 0 \, .
\label{eq:appI2}
\end{equation}
As a consequence, from the Boltzmann
equation set, we obtain
\begin{equation}
\sum_j \int \frac{d^3 p}{(2 \pi
\hbar)^3} \, \chi_j \, \left({ \frac{\partial f_j}{\partial t} }
+ \frac{\bf p}{m_j} \cdot \frac{\partial{f_j}}{\partial{\bf r}} +
{\bf F}_j \cdot \frac{\partial{f_j}}{\partial{\bf p}}\right) = 0
\, .
\label{eq:appI3}
\end{equation}
After a partial integration, we get
from the above
\begin{equation}
\frac{\partial}{\partial t} \left( n
\underline{\chi} \right) + \frac{\partial }{\partial {\bf r}}
\cdot \left( n \, \underline{\frac{\bf p}{m}\, \chi} \right) - n
\, \underline{{\bf F} \cdot \frac{\partial \chi}{\partial {\bf
p}}} = 0 \, ,
\label{eq:appI4}
\end{equation}
where the averages are defined
with
\begin{equation}
n \, \underline{\chi} = \sum_j \int \frac{d^3 p}{(2 \pi
\hbar)^3} \, \chi_j \, f_j({\bf p},{\bf r},t) \, .
\label{eq:appI5}
\end{equation}

Substituting for $\chi_j$ the conserved quantities
($\chi_j ({\bf p}) = \delta_{ij}$, ${\bf p}$ or $p^2/2m_j$),
we get the respective continuity equations:
\begin{subequations}
\label{eq:appI6}
\begin{eqnarray}
&& \frac{\partial n_i}{\partial t} + \frac{\partial }{\partial {\bf r}} \cdot
\left( n_i \, \underline{\bf v}_i \right) = 0 \, \\
&& \frac{\partial}{\partial t} \left( \rho \, \underline{\bf v} \right)
+ \frac{\partial}{\partial {\bf r}} \cdot \overline{\overline{p}}
- n_1 \, {\bf F}_1 - n_2 \, {\bf F}_2 = 0 \, , \\
&& \frac{\partial}{\partial t} \left( n \, \underline{e}\right)
+\frac{\partial}{\partial {\bf r}} \cdot {\bf q} - n_1 \,
\underline{\bf v}_1 \cdot {\bf F}_1 - n_2 \, \underline{\bf v}_2
\cdot {\bf F}_2 =0 \, .
\end{eqnarray}
\end{subequations}
Here, we made yet no use of the local frame.

The local frame is useful when wants to make use of the assumption of local equilibrium
that imposes restrictions on local quantities.
On representing the average velocities as $\underline{\bf v}_i
= \underline{\bf V}_i + \underline{\bf v}$ in the equations above, we obtain
the following set,
\begin{subequations}
\label{eq:appI8}
\begin{eqnarray}
&& \frac{\partial n_i}{\partial t} + \frac{\partial}{\partial {\bf r}} \cdot \left( n_i \,
\underline{\bf v} \right) + \frac{\partial}{\partial {\bf r}} \cdot \left( n_i \,
\underline{\bf V}_i \right) = 0 \, , \\
&& \frac{\partial \rho}{\partial t} + \frac{\partial}{\partial {\bf r}} \cdot \left( \rho \,
\underline{\bf v} \right) = 0 \, , \\
&& \frac{\partial }{\partial t} \left( \rho {\bf v} \right) +
\frac{\partial}{\partial {\bf r}} \cdot \left(\rho \, {\bf v \, v} \right)
+ \frac{\partial}{\partial {\bf r}} \cdot \overline{\overline{P}}
- n_1 \, {\bf F}_1 - n_2 \, {\bf F}_2 = 0 \, , \\
&& \frac{\partial}{\partial t} \left(n \, \underline{E} \right) +
\frac{\partial}{\partial {\bf r}} \cdot \left(n \, \underline{E} \, {\bf v} \right)
+ \overline{\overline{P}} : \overline{\overline{\frac{\partial}{\partial {\bf r}} \, {\bf v}}}
+ \frac{\partial}{\partial {\bf r}} \cdot {\bf Q} - n_1 \, \underline{\bf V}_1 \, {\bf F}_1
- n_2 \, \underline{\bf V}_2 \, {\bf F}_2 = 0 \, .
\end{eqnarray}
\end{subequations}
The equation for mass density in the set above follows from combining the equations for
particle densities.

The above equations significantly simplify when the assumption of a strict
local equilibrium
is imposed.  Under that assumption, the local species velocities and the heat flow
vanish, ${\bf V}_i = 0$ and ${\bf Q} = 0$, and the kinetic pressure tensor becomes diagonal,
$\overline{\overline{P}} = \frac{2}{3} \, n \, \underline{E}  \, \overline{\overline{1}}$.
The equations reduce then to the Euler set
\begin{subequations}
\label{eq:Eulers}
\begin{eqnarray}
\label{eq:Euler}
&& \frac{\partial n_i}{\partial t} + \frac{\partial}{\partial {\bf r}} \cdot \left( n_i \,
\underline{\bf v} \right) = 0 \, , \\
&& \frac{\partial }{\partial t} \left( \rho {\bf v} \right) +
\frac{\partial}{\partial {\bf r}} \cdot \left(\rho \, \overline{\overline{\bf v \, v}} \right)
+ \frac{2}{3} \, \frac{\partial (n\, \underline{E})}{\partial {\bf r}}
- n_1 \, {\bf F}_1 - n_2 \, {\bf F}_2 = 0 \, , \\
&& \frac{\partial}{\partial t} \left(n \, \underline{E} \right) +
{\bf v} \cdot \frac{\partial}{\partial {\bf r}} \left( n \, \underline{E} \right)
+ \frac{5}{3} \, n \, \underline{E} \, \frac{\partial}{\partial {\bf r}} \cdot {\bf v}
= 0 \, .
\end{eqnarray}
\end{subequations}

\section{Space-Time Derivatives for an Ideal Fluid}
\label{FermiSysEOS}

In an ideal fluid, all local quantities can be expressed in terms of the local temperature $T$
and the local kinetic chemical potential $\mu_i$.  If we consider changes of the densities $n_i$
or of the local kinetic energies $\underline{E}_i$
with respect to a parameter $x$ representing
some spatial coordinate or time, or their combination, we find
\begin{eqnarray}
\nonumber
 \frac{\partial n_i}{\partial x} & = & \xi_i \, T \, \frac{\partial
\alpha_i}{ \partial x}
+ \frac{3}{2} \, n_i \, \frac{\partial \beta}{ \partial x}\, ,  \\[.5ex]
 \frac{\partial  (n_i \, \underline{E}_i)}{\partial x} & = &
\frac{3}{2} \, n_i \, T \frac{\partial \alpha_i}{\partial x} +
\frac{5}{2} \, n_i \, \underline{E}_i \, \frac{\partial \beta}
{\partial x} \, ,
 \end{eqnarray}
where $\alpha_i = \mu_i / T$, $\beta = \log{T}$ and
$\xi_i=(\partial n_i / \partial \mu_i)_T$. With the trace
derivative defined as \begin{equation} \nonumber \frac{d}{dt} =
\frac{\partial}{\partial t} + {\bf v} \cdot
\frac{\partial}{\partial {\bf r}} \, , \end{equation} a particular version
of the above relations is
\begin{eqnarray}
\nonumber
 \frac{d n_i}{d t} &=& \xi_i \, T \, \frac{d
\alpha_i}{dt}
+ \frac{3}{2} \, n_i \, \frac{d \beta}{d t}\, ,  \\[.5ex]
\frac{d  (n_i \, \underline{E}_i)}{d t} &=&
\frac{3}{2} \, n_i \, T \frac{d \alpha_i}{d t} +
\frac{5}{2} \, n_i \, \underline{E}_i \, \frac{d \beta}
{d t} \, .
\label{eq:tracedv}
\end{eqnarray}

A combination of the above trace-derivative relations with the
Euler equations from Appendix \ref{continueE} yields the following
simple results,
\begin{subequations}
\label{eq:appc3}
\begin{eqnarray}
\frac{d \alpha_i}{dt}&=&0 \, , \label{eq:appc3a}\\
\frac{d \beta}{dt}&=& - \frac{2}{3} \, \frac{\partial}{\partial
{\bf r}} \cdot {\bf v} \, ,
\label{eq:appc3b}
\end{eqnarray}
\end{subequations}
the consistency of which with (\ref{eq:tracedv}) and
(\ref{eq:Eulers}) is easy to verify.
The results (\ref{eq:appc3}) express basic features
of the isentropic ideal-fluid evolution of
a mixture.
The entropy per particle in species $i$ depends only on $\alpha_i$, while the ratio
of the densities of species $n_1/n_2$ depends both on $\alpha_1$ and $\alpha_2$.  The
conservation of $\alpha_i$ for both species is equivalent to the conservation of entropy per
particle and of relative concentration.  Finally, the density for species $i$ is proportional
to $T^{3/2}$ multiplying a function of $\alpha_i$, which is equivalent to the second of
the results above, given the continuity equation for species and the conservation of $\alpha_i$.

\section{Variable Transformation}
\label{variable-transformation}

The driving forces for diffusion are naturally expressed in
terms of the gradients of temperature and of chemical
potential difference per unit mass $\mu_{12}^t$.  However,
given the typical constraints on systems, it can be convenient
to express the chemical potential in terms of other quantities,
that are easier to assess or control, such as the differential
concentration $\delta$, temperature $T$ and net pressure $P^t$.
A transformation of the variables for the driving forces
has been employed, at a formal level, in Sec.\ \ref{firstorder}.
Here, we show, though, how the transformation can be done in practice for
the interaction energy per particle specified in terms of
the particle density $n$ and concentration $\delta$, $E^v=E^v(n,\delta)$.
With the nuclear application in mind, we limit ourselves to
the case of $m_1=m_2=m$.

The transformation can exploit straightforward relations between
different differentials.  One of those to exploit is the Gibbs-Duhem relation
\begin{eqnarray}
dP^t= n_1 \, d\mu_1^t + n_2 \, d\mu_2^t + n\, s\, dT
=n \, d\mu^t + \frac{m \, n\, \delta}{2} \, d\mu_{12}^t + n \, s \, dT \, .
\end{eqnarray}
Here, $s$ is the entropy per particle and $\mu^t = (\mu_1^t + \mu_2^t)/2$ is
the median chemical potential.  Two other relations stem
from the differentiations of equilibrium particle distributions,
already utilized in Appendix \ref{FermiSysEOS},
\begin{equation}
dn_i  =  \xi_i \, d\mu_i + \frac{\frac{3}{2}n_i - \xi_i \, \mu_i}{T} \, dT
\equiv \xi_i \, d\mu_i + \left( \frac{\partial n_i}{\partial T} \right)_{\mu_i} \, dT \, .
\end{equation}
With $\mu_i^v = {\partial (n \, E^v)}/{\partial n_i}$, on adding and subtracting
the two ($i=1,2$) relations side by side, we find
\begin{eqnarray}
\nonumber
\hspace*{-1em} dn & = & \left(\xi_1 + \xi_2\right)
\left[ d\mu^t - \left(\frac{\partial \mu^v}{\partial n} \right)_\delta \,
dn - \left(\frac{\partial \mu^v}{\partial \delta} \right)_n \, d\delta \right]
 + \frac{m}{2} \left( \xi_1 - \xi_2 \right) \\ & & \times
\left[ d\mu_{12}^t - \left(\frac{\partial \mu_{12}^v}{\partial n} \right)_\delta \,
dn - \left(\frac{\partial \mu_{12}^v}{\partial \delta} \right)_n \, d\delta \right]
+ \left[ \left( \frac{\partial n_1}{\partial T} \right)_{\mu_1}
+ \left( \frac{\partial n_2}{\partial T} \right)_{\mu_2}\right] \, dT \,
\label{eq:G1}
\end{eqnarray}
and
\begin{eqnarray}
\nonumber
\delta \, dn & + & n \, d\delta  =
\left(\xi_1 - \xi_2\right)
\left[ d\mu^t - \left(\frac{\partial \mu^v}{\partial n} \right)_\delta \,
dn - \left(\frac{\partial \mu^v}{\partial \delta} \right)_n \, d\delta \right]
 + \frac{m}{2} \left( \xi_1 + \xi_2 \right) \\  & & \times
\left[ d\mu_{12}^t - \left(\frac{\partial \mu_{12}^v}{\partial n} \right)_\delta \,
dn - \left(\frac{\partial \mu_{12}^v}{\partial \delta} \right)_n \, d\delta \right]
+ \left[ \left( \frac{\partial n_1}{\partial T} \right)_{\mu_1}
- \left( \frac{\partial n_2}{\partial T} \right)_{\mu_2}\right] \, dT \, .
\label{eq:G2}
\end{eqnarray}
Those two equations have the structure
\begin{equation}
G_{kn} \, dn   = G_{k \mu} \, d\mu^t + G_{k d} \, d\mu_{12}^t + G_{k \delta} \, d\delta
+ G_{kT} \, dT \, ,
\end{equation}
where $k=1,2$ and where the coefficients $G$ can be worked
out from (\ref{eq:G1}) and (\ref{eq:G2}).  On multiplying
the sides of the first ($k=1$) equation by $G_{2n}$
and the sides of the second ($k=2$) equation by $G_{1n}$
and on subtracting
the equations side by side, we can eliminate the $dn$ differential obtaining
\begin{eqnarray}
\nonumber
0 &= & (G_{2n} \, G_{1 \mu} - G_{1n} \, G_{2 \mu}) \, d\mu^t + (G_{2n} \, G_{1 d} - G_{1n} \, G_{2 d}) \, d\mu_{12}^t
\\ \nonumber &&
+ (G_{2n} \, G_{1 \delta} - G_{1n} \, G_{2 \delta}) \, d \delta + (G_{2n} \, G_{1 T} - G_{1n} \, G_{2 T}) \, dT
\\
& \equiv & R_{\mu} \, d\mu^t + R_d \, d\mu_{12}^t + R_\delta \, d \delta + R_T \, dT \, .
\end{eqnarray}
On eliminating next the $d\mu_t$ differential using the Gibbs-Duhem
relation, we find
\begin{subequations}
\begin{eqnarray}
\Pi_{12}^P & = & \left(\frac{\partial \mu_{12}^t}{\partial P^t}\right)_{T,\delta}
=  \frac{R_\mu}{n \, \left(R_\mu \, \frac{m \, \delta}{2} - R_d \right)} \, , \\
\Pi_{12}^T &=& \left(\frac{\partial \mu_{12}^t}{\partial T}\right)_{P^t,\delta}
= \frac{R_\mu \, s- R_T}{R_d - R_\mu \, \frac{m \, \delta}{2}} \, , \\
\Pi_{12}^\delta & = & \left(\frac{\partial \mu_{12}^t}{\partial \delta}\right)_{P^t,T}
=  \frac{R_\delta}{R_\mu \, \frac{m \, \delta}{2} - R_d } \, .
\end{eqnarray}
\end{subequations}

\section{Brace Algebra}
\label{bracket-algebra}

The brace products are employed in finding the transport coefficients
within linear approximation to the Boltzmann equation.  The brace product
of two scalar quantities $A$ and $B$ associated with the colliding particles
is defined as
\begin{eqnarray}
\nonumber
\lbrace A,B \rbrace & = & \int \frac{d^3p}{(2\pi)^3} \, A_1 \, I_{11}(B)
+ \int \frac{d^3p}{(2\pi)^3} \, A_1 \, I_{12}(B) \\ \nonumber
&& + \int \frac{d^3p}{(2\pi)^3} \, A_2 \, I_{21}(B)
+ \int \frac{d^3p}{(2\pi)^3} \, A_2 \, I_{22}(B) \\
& = & \left[A,B\right]_{11} + \left[A,B\right]_{12} + \left[A,B\right]_{22}
\, ,
\label{curlyCD}
\end{eqnarray}
where, in the last step, we have broken the brace product
into square-bracket products representing contributions from collisions
within species $1$, from collisions between species $1$ and $2$ and from collisions
within species $2$, respectively.

We will first show that the square-bracket product is symmetric.
Thus, we have explicitly
\begin{eqnarray}
\nonumber
\left[A,B\right]_{ii} & = & \frac{1}{2} \int \frac{d^3p_a}{(2\pi)^3}
\, \frac{d^3p_b}{(2\pi)^3} \, d\Omega \, v^* \,
\left( \frac{d\sigma_{ii}}{d\Omega} \right) \, f_{ia}^{(0)} \, f_{ib}^{(0)} \,
{\tilde{f}_{ia}^{(0)\prime}} \, {\tilde{f}_{ib}^{(0)\prime}} \, \\[.5ex] \nonumber
&& \times A_{ia} \, ( B_{ia}+B_{ib} - B_{ia}' - B_{ib}') \\[.5ex] \nonumber
& = & \frac{1}{8} \int \frac{d^3p_a}{(2\pi)^3}
\, \frac{d^3p_b}{(2\pi)^3} \, d\Omega \, v^* \,
\left( \frac{d\sigma_{ii}}{d\Omega} \right) \, f_{ia}^{(0)} \, f_{ib}^{(0)} \,
{\tilde{f}_{ia}^{(0)\prime}} \, {\tilde{f}_{ib}^{(0)\prime}} \, \\[.5ex]
&& \times (A_{ia} + A_{ib} - A_{ia}' - A_{ib}') \, ( B_{ia}+B_{ib} - B_{ia}' - B_{ib}') \, ,
\label{squarebracket}
\end{eqnarray}
where, to get the last result, we have first utilized an interchange of the particles
in the initial state of a collision and then an interchange
of the initial and final states within
a collision.  It is apparent that the r.h.s.\ of (\ref{squarebracket}) is
symmetric under the interchange of $A$ and $B$.  Moreover, we can see
that a square bracket for $B=A$, $[A,A]_{ii}$, is nonnegative and that it vanishes
only when $A$ is conserved in collisions.

We next consider the contribution from collisions between different species,
\begin{eqnarray}
\nonumber
\left[A,B\right]_{12} & = &  \int \frac{d^3p_1}{(2\pi)^3}
\, \frac{d^3p_2}{(2\pi)^3} \, d\Omega \, v^* \,
\left( \frac{d\sigma_{12}}{d\Omega} \right) \, f_{1}^{(0)} \, f_{2}^{(0)} \,
{\tilde{f}_{1}^{(0)\prime}} \, {\tilde{f}_{2}^{(0)\prime}} \, \\[.5ex] \nonumber
&& \times (A_{1}+A_{2}) \, ( B_{1}+B_{2} - B_{1}' - B_{2}') \\[.5ex] \nonumber
& = & \frac{1}{2} \int \frac{d^3p_1}{(2\pi)^3}
\, \frac{d^3p_2}{(2\pi)^3} \, d\Omega \, v^* \,
\left( \frac{d\sigma_{12}}{d\Omega} \right) \, f_{1}^{(0)} \, f_{2}^{(0)} \,
{\tilde{f}_{1}^{(0)\prime}} \, {\tilde{f}_{2}^{(0)\prime}} \, \\[.5ex]
&& \times (A_{1} + A_{2} - A_{1}' - A_{2}') \, ( B_{1}+B_{2} - B_{1}' - B_{2}') \, .
\label{squarebracketsym}
\end{eqnarray}
Here, we again utilized an interchange between the initial and final states
and we again observe a symmetry between $A$ and $B$ on the r.h.s.
Thus, indeed, all square
brackets are symmetric.  Moreover,
for $B=A$, we see that $[A,A]_{12} \ge 0$ and that the zero is only reached if $A$ is
conserved.

Combining the results, we find that the brace product
(\ref{curlyCD}) is symmetric. Moreover, we find that the brace product of
quantity $A$ with itself is nonnegative, $\lbrace A, A \rbrace \ge
0$, and vanishes only when $A$ is conserved.  As the brace product
has features of a pseudo-scalar product, a version of the
Cauchy-Schwarz-Buniakowsky (CSB) inequality \cite{Gradshteyn}
holds, \begin{equation} \lbrace A, A \rbrace \, \lbrace B, B \rbrace \ge
\left( \lbrace A, B \rbrace \right)^2 \, . \end{equation}

All the results from this Appendix remain valid, in an obvious manner, when
the brace product (\ref{curlyCD}) is generalized
to the pairs of tensors of the same rank
associated with the particles, when requiring that the tensor indices are
convoluted between the two tensors in the brace, as e.g.\ in (\ref{eq:velocitydiff}).
The positive definite nature of the brace product is important in ensuring that
expressions for transport coefficients, obtained in the paper, yield positive values for the
coefficients that in this case represent a stable system.

\newpage

\begin{table}
\caption{
Diffusion coefficient $D_I$
obtained within different orders of calculation,
using
experimental np cross sections,
at sample
densities $n$ and temperatures $T$
in symmetric nuclear matter,
for species-independent mean fields.
The numerical errors of the results on $D_I$ are indicated in parenthesis
for the least-significant digits.  The last two columns,
separated by the '$\pm$' sign, give, respectively, the relative change in the result for the
highest calculated order compared to the first order and the error
for that change.
\\[1ex]}
\begin{tabular}{cccccrcl}
\colrule
n & T& & $D_I$&  &\multicolumn{3}{c}{Relative}\\
 & & $1^{\mbox{st}}$ order &  $2^{\mbox{nd}}$ order & $3^{\mbox{rd}}$ order & \multicolumn{3}{c}{Change} \\
 fm$^{-3}$ & MeV &  & fm$\, c$ &   & \multicolumn{3}{c}{$\%$} \\
\colrule
 0.016 & 10 & 0.29949(15) & 0.3055(12) & & 2.0 & $\pm$ & 0.4\\
 0.016 & 60 & 2.3891(18)  & 2.390(14)  & & 0.0 & $\pm$ & 0.6\\
 0.16  & 10 & 0.27964(21) & 0.2800(29) & 0.2809(25) & 0.5 & $\pm$ & 0.9\\
 0.16  & 60 & 0.29591(24) & 0.2965(19) & & 0.2 & $\pm$ & 0.7 \\
 0.32  & 10 & 0.4446(15)  & 0.4465(26) & & 0.4 & $\pm$ & 0.7 \\
 0.32  & 60 & 0.18187(15) & 0.1827(13) & & 0.5 & $\pm$ & 0.7 \\
\colrule
\end{tabular}
\label{TablehigherOrder}
\end{table}

\begin{thebibliography}{99}

\bibitem{lat01}
J.\ M.\ Lattimer and M.\ Prakash, Astrophys.\ J.\ 550, 426 (2001).


\bibitem{Rami99} F.\ Rami et al.,
Phys.\ Rev.\ Lett.\ {\bf 84}, 1120 (2000).


\bibitem{Bertsch} G.\ F.\ Bertsch, S.\ Das Gupta,
Phys.\ Rep.\ {\bf 160}, 189 (1988).

\bibitem{Aichelin} J.\ Aichelin,
Phys.\ Rep.\ {\bf 202}, 233 (1991).


\bibitem{Tomonaga}
S.\ Tomonaga, Z.\ Phys.\ 110, 573 (1938).

\bibitem{gal79}
V.\ M.\ Galitskii et al., Yad.\ Fiz.\ 30, 778 (1979) [Sov.\ J.\ Nucl.\
Phys.\ 30, 401 (1979).]

\bibitem{PD84} P.\ Danielewicz,
Phys.\ Lett.\ {\bf 146 B}, 168 (1984).

\bibitem{hak92}
R.\ Hakim, L.\ Mornas, P.\ Peter, and H.\ D.\ Sivak, Phys.\ Rev.\ D 46,
4603 (1992).

\bibitem{hak93}
R.\ Hakim and L.\ Mornas,
Phys.\ Rev.\ C 47, 2846 (1993).



\bibitem{dan02}
P.\ Danielewicz,
Acta Phys.\ Polonica B 33, 45 (2002).


\bibitem{Chapman} S.\ Chapman and T.\ G.\ Cowling,
{\it The Mathematical Theory of Non-Uniform Gases}
(Cambridge, New York, 1964).


\bibitem{ueh33}
E.\ A.\ Uehling and G.\ E.\ Uhlenbeck,
Phys.\ Rev.\ 43, 552 (1933).

\bibitem{ueh34}
E.\ A.\ Uehling, Phys.\ Rev.\ 46, 917 (1934).

\bibitem{hel39}
E.\ J.\ Hellund and E.\ A.\ Uehling,
Phys.\ Rev.\ 56, 818 (1939).


\bibitem{Liboff} R.\ L.\ Liboff,
{\it Kinetic Theory, Classical, Quantum, and Relativistic Descriptions}
(Prentice Hall, Englewood Cliffs, New Jersey, 1990).



\bibitem{Landau} L.\ D.\ Landau and E.\ M.\ Lifshitz,
{\it Fluid Mechanics}, Vol.\ 6 of {\em Course
of Theoretical Physics} (Addison-Wesley, Reading, 1959).


\bibitem{Groot} S.\ R.\ De Groot and P.\ Mazur,
{\it Non-Equilibrium Thermodynamics}
(North-Holland, Amsterdam, 1962).


\bibitem{Baym} G.\ Baym and C.\ Pethick,
{\it Landau Fermi-Liquid Theory}
(John Wiley \& Sons, New York, 1991).


\bibitem{Toro}
V.\ Greco, M.\ Colonna, M.\ Di Toro and F.\ Matera,
Phys.\ Rev.\ C67, 015203(2003).


\bibitem{yen94}
S.\ J.\ Yennello et al., Phys.\ Lett.\ B 321, 15 (1994);
H.\ Johnston et al., Phys.\ Lett.\ B 371, 186 (1996);
H.\ Johnston et al., Phys.\ Rev.\ C 56, 1972 (1997).



\bibitem{Gradshteyn} I.\ S.\ Gradshteyn and I.\ M.\ Ryzhik,
{\it Table of Integrals, Series, and Products}
(Academic Press, New York, 1979).


\end{thebibliography}
\end{document}